\documentclass[11pt, draftcls,onecolumn]{IEEEtran} 

\newcommand{\prob}{\mathbb{P}}

\def\prob#1{ {\text{Pr}} \left \{  #1 \right \}}
\def\mtx#1{{\text{#1}}}

\newcommand{\XX}{V}

\newtheorem{proposition}{Proposition}
\newtheorem{theorem}{Theorem}
\newtheorem{example}{Example}
\newtheorem{remark}{Remark}

\newtheorem{definition}{Definition}

\newtheorem{lemma}{Lemma}

\newcommand{\recset}{ {\cal R} }
\newcommand{\retset}{ {\cal T} }

\usepackage{epsfig,graphics,psfrag,subfigure}
\usepackage{amsmath,amssymb}

\begin{document}
\title{On the Information Rates of the Plenoptic Function}
%
\author {Arthur~L.~da~Cunha,
         Minh~N.~Do,~\IEEEmembership{Member,~IEEE,}
         and~Martin~Vetterli~\IEEEmembership{Fellow,~IEEE}
\thanks{A. da Cunha and Minh Do are with the Department of Electrical and
Computer Engineering and the Coordinated Science Laboratory,
University of Illinois at Urbana-Champaign, Urbana IL. (email:
cunhada@uiuc.edu, minhdo@uiuc.edu)}
    \thanks{M. Vetterli is with Laboratory for Audiovisual Communications, Ecole Polytechnique F\'{e}d\'{e}rale de Lausanne (EPFL), and
    Dept. of EECS, UC Berkeley, CA, USA. (email:
    martin.vetterli@epfl.ch)} \thanks{This work was presented in part at the Data Compression Conference (DCC), Snowbird, UT, 2007
    \cite{CunhaDoV:07}.}}

\markboth{IEEE Transactions on Information Theory}{Shell
\MakeLowercase{\textit{et al.}}: Bare Demo of IEEEtran.cls for
Journals} \maketitle
\begin{abstract}
The {\it plenoptic function} describes
the visual information available to an observer at any point in
space and time. Samples of the plenoptic function (POF) are seen
in video and in general visual content (images, mosaics, panoramic scenes, etc), and represent large
amounts of information. In this paper we propose a stochastic
model to study the compression limits of a simplified version of the plenoptic function.
In the proposed framework, we isolate the two fundamental sources
of information in the POF: the one representing the camera motion
and the other representing the information complexity of the
``reality'' being acquired and transmitted. The sources of
information are combined, generating a stochastic process that we
  study in detail.

We first propose a model for ensembles of realities that do not change over time.
The proposed model is
simple in that it enables us to derive precise coding bounds in
the information-theoretic sense that are sharp in a number of cases of
practical interest. For this simple case of static realities and camera motion,
our results indicate that coding practice is in accordance with optimal coding from an
information-theoretic standpoint.

The model is further extended to account for visual realities that
change over time.
We derive bounds on the lossless
and lossy information rates for this dynamic reality model,
stating conditions under which the bounds are tight. Examples with synthetic sources suggest
that within our proposed model, common hybrid coding using motion/displacement
estimation with DPCM performs considerably suboptimally relative
to the true rate-distortion bound.

\end{abstract}



\IEEEpeerreviewmaketitle

\newpage

\section{Introduction} \label{sec:intro}

\subsection{Background}
    {C}{onsider} a moving camera that takes sample
    snapshots of an environment over time. The samples are to be coded
    for transmission or storage. Because the movements of the
    camera are small relative to the scene, there are large
    correlations among multiple acquisitions.

    Examples of such scenarios include video compression and the
    compression of light-fields. More generally, the compression problem
    in these examples can be seen as representing and compressing
    samples of the {\it plenoptic function} \cite{AdelsonBerg:91}.
     The 7-D plenoptic function (POF) describes the light intensity passing
    through every viewpoint, in every direction, for all times, and for
    every wavelength. Thus, the samples of the plenoptic function can
     be used to reconstruct a view of reality at the decoder. The POF
    is usually denoted by $POF(x,y,z,\phi,\varphi,t, \lambda)$, where
    $(x,y,z)$ represents a point in 3-D space, $(\phi,\varphi)$
    characterizes the direction of the light rays, $t$ denotes
    time, and $\lambda$ denotes the wavelength of the light rays. The
    POF is usually parametrized in order to reduce its number of
    dimensions. This is common in image based rendering
    \cite{ForsythP:02bk,ZhangChen:04}. Examples of POF
    parameterizations include digital video, the lightfield and
    lumigraph \cite{LevoyHanrahan:96,GortlerGr:96}, concentric mosaics
    \cite{HeShumConcM:99}, and the surface plenoptic function
    \cite{ZhangChenSPF:03}. Regardless of the parametrization, due to
    the large size of the data set, compression is essential.
    
    In this work, we consider the plenoptic function in terms of a spatial position and a time dimension. Thus, our initial setup is that of $POF(x,y,z,t)$. We also assume that we do not have information on the constituents dimensions, but rather we are given a sampled plenoptic function that needs to be compressed. A typical scenario involves a camera traversing
    the domain of the POF and acquiring its samples to be compressed and then
    stored for later rendering. The
    information to be compressed is thus $POF(W(t),t)$ where the
    trajectory $W(t)$ collectively represents  a sequence of positions
    and angles where light rays are  acquired. In such a context, it is
    crucial to know the compression limits and how the parameters
    involved influence such limits. This then provides a benchmark to
    assess the performance of compression schemes for such data sets.


%
%

\subsection{Prior art}

    The practical aspects of compressing video and other examples of the plenoptic function
    have been studied extensively (see e.g.,
    \cite{Telkap:95bk,ZhangChenSPF:03,WangZhang:02,Woods:06}, and references
    therein). But very little has been done in terms of rate-distortion analysis and
    addressing the general question of how many bits are needed to code
    such a source. Due to the complexity inherent to visual data, the source is
    difficult to model statistically. As a result, precise information
    rates are difficult to obtain. Several statistical models have
    been proposed to analyze video sources \cite{ForchheimerLi:96,Girod:99,CHerley:06,Vasconcelos:00}. Often, one obtains the
    rate-distortion behavior resulting from a particular coding method,
    such as the hybrid coder used in video. For instance, the work in
    \cite{Girod:87} analyzes the rate-distortion performance of hybrid
    coders using a Gauss-Markov model for the video sequence as well as
    for the prediction error that is transmitted as side information. A
    similar rate-distortion analysis for light-field compression is done
    in \cite{RamanathanG:05}. Such models are interesting but they work
    with the assumption of predictive coding from the start, and thus they do not reveal the \emph{intrinsic} information 
    rate of the visual source. 
    
    The compression of the POF is also studied
    in \cite{GehrigD:04}, but in
    a distributed setting. Using
    piece-wise smooth models, the authors derive operational
    rate-distortion bounds based on a parametric sampling model. Another scenario of POF coding is studied in \cite{Kauff:05}.

\begin{figure}[hb]
\begin{footnotesize}\centering
\psfrag{VIEW OF}{\tiny VIEW OF} \psfrag{REALITY}{\tiny REALITY}
\psfrag{X}{\tiny $\XX$} \psfrag{Xh}{\tiny $\hat \XX$}
\psfrag{ENCODER}{\tiny ENCODER} \psfrag{DECODER}{\tiny DECODER}
\psfrag{MEMORY}{\tiny \ \ \ \ WITH } \psfrag{M}{\tiny
\hspace{-.7cm} MEMORY $M$} \psfrag{CHANNEL}{\tiny CHANNEL}
\psfrag{RATE}{\tiny \hspace{-.4cm} RATE $R$}
\psfrag{R}{\footnotesize ``Reality''}
\includegraphics[width=15.4cm ]{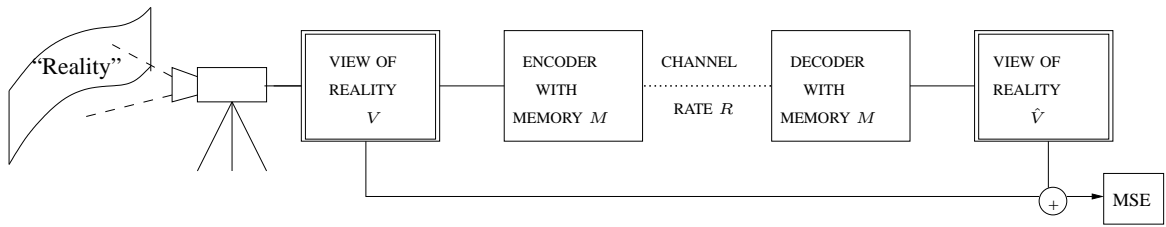}
\caption{The problem under consideration. There is a world and a
camera that produces a ``view of reality'' that needs to be coded
with finite or infinite memory. \label{Fig:reality}}
\end{footnotesize}
\end{figure}

 \subsection{Paper contributions}

    The general problem can be posed as shown in Figure
    \ref{Fig:reality}. There is a
    physical world or ``reality'' (e.g., scenes, objects, moving
    objects), and a camera that generates a ``view of reality'' $\XX$. This
    ``view of reality'' (e.g., a video sequence) is coded with a source
    coder with memory $M$  giving an average rate of
    $R$ bits per sample.  This bitstream is decoded with a decoder with memory $M$
    to reconstruct a view of reality $\hat \XX$ close to the original one in the MSE
    sense. We refer to memory and rate in a loose sense. Precise
    definitions of memory and rate are given in Section
    \ref{subsec:Video_Coding}.

    In this paper we propose a simplified
    stochastic model for the plenoptic function that bears the essential elements of the
    general case. We take the viewpoint that  video can be seen
    as a 3-D slice of the POF. Our approach is to come up with
    a statistical model for video data generation, and within that
    model establish information rate bounds. We first propose
    a model in which the background scene is drawn randomly at a prior time, but otherwise does not change as time progresses. Within this ``static reality'' model
    we develop information rates for the lossless and lossy cases. Furthermore, we compute the conditional information
    rate that provides a coding limit when memory resources are constrained. We then extend the model to account for background scene changes. We then propose a ``dynamic reality'' that is based
        on a Markov random field. We compute bounds on the information rates.
        For the Gaussian case, we compute lower and upper bounds that are tight in the high SNR regime. Examples
        validating our theoretical findings are presented.
%
	
	The models proposed and studied in this paper make several assumptions to make the problem mathematically tractable. Our goal here is to make assumptions that simplify the problem but still keep the main elements of the general problem of compressing data from a moving camera. While the resulting models are not a perfect depiction of reality we believe they have merit as they provide a framework to investigate such processes. What is more, our assumptions allow us to derive coding bounds that to the best of our knowledge are unknown, even in the case of our very simplified models.   

        The paper is organized as follows. Section \ref{SecII:ProblemSetup} sets up the problem and introduces
        notation. The video coding problem is treated in Sections \ref{SecIII:lossless} and \ref{SecIV:innovation}.
        We present results for the static reality case in Section \ref{SecIII:lossless}, and treat the dynamic case
        in Section \ref{SecIV:innovation}.  Concluding remarks are made in Section
        \ref{Sec:Conc}.
\section{Definitions and Problem Setup}\label{SecII:ProblemSetup}
    \subsection{Simplified model}
    We describe a simplified model for the process displayed in Figure
    \ref{Fig:reality}. Consider a camera moving according
    to a Bernoulli random walk. The random walk is defined as follows:

    \begin{definition}\label{DEF:rwalk}
    The Bernoulli random walk is the process $W = (W_t:t\in\mathbb{Z}^+)$ such
    that $\prob{W_0=0}=1$ and for $t\geq1$, $$W_t = \sum_{i=1}^tN_i,$$
    where $\{N_i\}$ are drawn i.i.d. from the set $\{-1,1\}$
    with probability distribution $Pr\{N_i = 1\} = p_W$.
    \end{definition}

    We assume without loss of generality that $p_W\leq  0.5$. Moreover, throughout the paper, the index $t$ is considered a discrete-valued variable. 

\begin{figure*}[htb]
        \centering \psfrag{Sn}{\tiny $W_t$} \psfrag{d}{\hspace{-0.2cm} $\cdots$}
        \psfrag{X1}{  \tiny $X_{0}$  } \psfrag{X2}{\ \tiny $X_{1}$}
        \psfrag{X3}{\tiny $X_{2}$} \psfrag{X4}{\tiny $X_{3}$  }
        \psfrag{Camera}{ { \tiny Camera Position } }
        \psfrag{Image(n)}{ \ \ \raisebox{.2cm}{\tiny Image(t) } }
        \psfrag{X}{ \hspace{-.2cm} \raisebox{.1cm}{\tiny $ \XX_0:=(X_{0},   X_{1}, X_{2}, \XX_3$) } }
        \subfigure[]{\includegraphics[width=6.5cm, height=4.5cm]{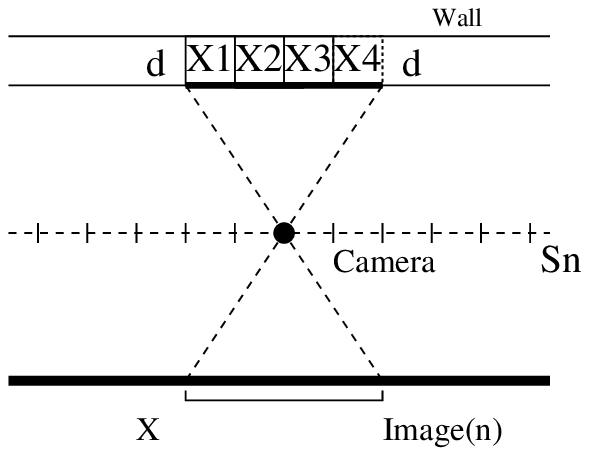}} \ \ \ \ \
        \psfrag{X0}{\tiny $\XX_0$} \psfrag{X1}{\tiny
        $\XX_1$} \psfrag{X2}{\tiny $\XX_2$}
        \psfrag{X3}{\tiny $\XX_3$} \psfrag{X4}{\tiny
        $\XX_4$} \psfrag{X5}{\tiny $\XX_5$}
        \psfrag{X6}{\tiny  $\XX_6$} \psfrag{X}{\tiny  Wall process $X$} \psfrag{W}{\tiny $W_t$} \psfrag{dots}{\small
        $\vdots$} \psfrag{ldots}{\small $\cdots$}
         \subfigure[]{\includegraphics[width=7.1cm, height=4.5cm]{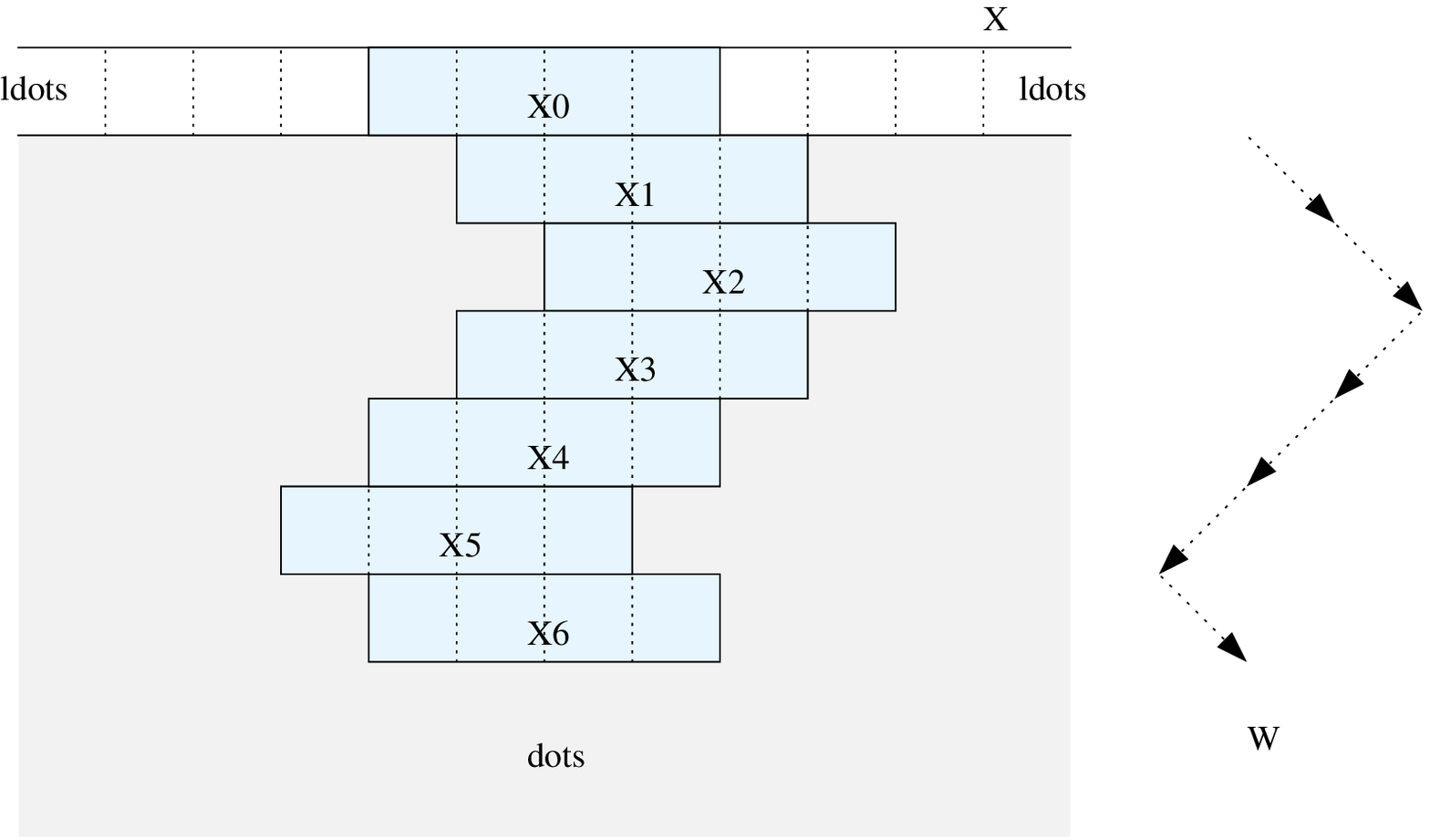}}
         \caption{A stochastic model for video. (a) Simplified model. (b) The
        resulting vector process $\XX$. Each sample of the vector process is
        a block of $L$ samples from the process $X$ taken at the
        position indicated by the random walk $W_t$. In the figure $L=4$.
        \label{Fig:vprocess}}
         \end{figure*}

    In front of the camera there is an infinite wall that represents a
    scene that is projected onto a screen in front of the camera path
    (i.e., we ignore occlusion). The wall is modelled as a 1-D strip
    ``painted" with an i.i.d. process $X= (X_n: n\in \mathbb{Z})$ that
    is {\it independent} of the random walk $W$. The process $X$
    follows some probability distribution $p_X$ drawing values from an alphabet
    ${\cal X}$. Here we focus on the rather unrealistic i.i.d.
    case due to its simplicity. Generalization to stationary process is left for future work.
    In the static case, the wall process $X$ is drawn at $t=0$.
    Figure  \ref{Fig:vprocess} (a) illustrates the proposed model.

    At each random walk time step, the camera sees a block of $L$
    samples from the infinite wall, where $L\geq 1$. This results in a vector
    process $\XX=(\XX_t:t\in \mathbb{Z}^+)$ indexed by the random walk
    positions, as defined below.

    \begin{definition} \label{Def:vprocess} Let $W$ be a random walk independent of $X$,
    and let $L$ be an integer greater than one. The vector process
    $\XX=(\XX_t:t\in\mathbb{Z}^+)$ is defined as
    \begin{equation}
    \XX_t:=(X_{W_t}, X_{W_{t}+1}, \cdots, X_{W_{t}+L-1}).\\
    \end{equation}
    \end{definition}

    The random walk is a simple stochastic model
    for an ensemble of camera movements. It includes {\it camera
    panning} as a special case, i.e., when $p_W=0$.
    The discrete displacements of the
    random walk thus neglect other effects such as zooming, rotation,
    and change of angle. 
        
    Notice that consecutive samples of the vector process, which are
    vectors of length $L$, have at least $L-1$ samples that are
    repeated. Furthermore,
    because the process $X$ is i.i.d.,
    it follows that the vector process $\XX$ is stationary and mean-ergodic.
    Figure  \ref{Fig:vprocess} (b) illustrates the vector process $\XX$.

    \subsection{The video coding problem }
    \label{subsec:Video_Coding}

    Given the vector process
    $\XX=(\XX_0,\XX_1,\cdots)$, the coding problem consists in finding
    an encoder/decoder pair that is able to describe and reproduce the
    process $\XX$ at the decoder using no more than $R$ bits per
    vector sample. The decoder reproduces the vector
    process $\hat \XX=(  \hat \XX_0,\hat \XX_1,\cdots)$ with some delay.
    The reproduction can be lossless or lossy with fidelity $D$. The
    encoder encodes each sample $\XX_t$  based on the observation of
    $M$ previous vector samples $\XX_{t-1}, \ldots, \XX_{t-M}$. Thus,
    $M$ is the memory of the encoder/decoder. Since encoding is done
    jointly, there is a delay incurred. The lossless and lossy information rates of the
    process $\XX$ provide the minimum rate needed to either
    perfectly reproduce the process $\XX$ at the decoder, or to
    reproduce it within distortion $D$, respectively. Note that the information rate (lossless or lossy) is
    usually only achievable at the expense of infinite memory and
    delay \cite{Berger:71}.

    \subsection{Properties of the random walk }\label{subsec:RWalk_Prop}
     The following notions are needed in this paper.

    \begin{definition} \label{DEF:RWALK1}Let $W$ be a random walk. The set of {\it
    recurrent} paths of length $t$ is the set
    $$\recset^t := \{ (W_0,W_1,\ldots, W_t): W_t=W_s  \text{ \ for some\  } s,\  0\leq s <t \}.$$
    If a path belongs to $\recset^t$, we call it a recurrent path. We call $\prob{{\cal R}^t}$ the {\it probability of
    recurrence} at step $t$.
    \end{definition}

    The probability of the complementary set
    $\prob{\overline{\recset^t}}$ is called the {\it first-passage}
    probability. When a site $W_t$ has not occurred before, we refer
    to it as a {\it new site}. A related quantity is the probability
    of return.

    \begin{definition} \label{DEF:RWALK2} Let $W$ be a random walk, and let $t>s\geq 0$. Consider the set
    $$\retset_s^t := \{  (W_0,W_1,\ldots, W_t): W_t = W_s \text{ but }  W_t \neq W_i, \text{ for any  $i$ such that $s<i<t$}  \}.$$
    We call $\prob{ \retset_s^t }$ the probability of return at
    step $t$ after step $s$.
    \end{definition}

    When $s=0$, we write $\retset^t$ for $\retset_0^t$. From
    Definitions \ref{DEF:RWALK1} and \ref{DEF:RWALK2} one can
    check that

    \begin{equation}\label{eq:disjoint_union} \recset^t =
    \bigcup_{i=1}^{ t } \retset^t_{t-i},
    \end{equation}
    \noindent where the union is a disjoint one. Furthermore, the sets $\retset_s^t$ are shift-invariant
    in the sense that \begin{equation} \label{eq:disjoint_union_prob}
    \prob{\retset_s^t} = \prob{\retset^{t-s}}.\end{equation}
    Combining (\ref{eq:disjoint_union}) and (\ref{eq:disjoint_union_prob}), we also have that
    \begin{equation}\label{eq:prob_union}
    \prob{\recset^t}=\sum_{i=0}^t \prob{\retset^t_{t-i}} = \sum_{i=0}^t \prob{\retset^i}.
    \end{equation}

    In addition to the above, for the case of
    the Bernoulli random walk we have the following \cite{RudnickGasp:04,WFeller:68bk}.

    \begin{lemma}\label{LEMMA:RWALK} For the Bernoulli random walk with $p_W \leq 1/2$, the following holds:
    \begin{description}
    \item[(i)] $\lim_{t \rightarrow \infty } \prob{\overline{\recset^t}} = 1-2p_W$.
    \item[(ii)] For $t>0$, $\prob{\retset^{2t-1}}=0$, and  $\prob{\retset^{2t}} = 2C_{t-1} \left ( (1-p_W)p_W
    \right)^t$, where $C_t := \frac{1}{t+1}{2t\choose t}$.
    \end{description}

    \end{lemma}

\section{Information Rates for a Static Reality}
\label{SecIII:lossless}

       \subsection{Lossless information rates for the discrete memoryless wall}

       Denote $\XX^t = (\XX_1,\ldots,\XX_t)$. We assume that
       $\XX_0$ is known to the decoder. Unless otherwise specified, we assume that $X$ takes values on
       a finite alphabet ${\cal X}$. We
       seek to quantify the entropy rate of $\XX$  \cite{CoverThomas:91}:
       \begin{equation} \label{eq:erate}
       H(\XX) =   \lim_{t\rightarrow \infty} \frac{1}{t}
       H\left (\XX^t \right ) = \lim_{t\rightarrow \infty} H(\XX_t|\XX^{t-1}).
       \end{equation}
%
       To characterize  $H(\XX)$, we
       describe intuitively an upper and a lower bound
       (resp. sufficient and necessary rates) that will be
       formalized in Theorem \ref{THM:LB_UB_V} below.
       For a sufficient rate, note that $\XX$ can be
       reproduced up to time $t$ when both the trajectory $W^t=(W_1,
       \ldots, W_t)$ and the samples of the wall occurring at the new
       sites of $W^t$ are available. When $t$ is large,
       this amounts to $H(W^t) = t H(p_W)$ bits for the
       trajectory, plus $t  \prob{\overline{\recset^t}} H(X) \approx t  (1-2p_W)H(X)$ for the new sites. So, a sufficient
       average rate is  $H(p_W) +
       (1-2p_W)H(X)$. Moreover, the complexity of $\XX$ is at least the total complexity of all visited new sites, and so
        $(1-2p_W) H(X)$ is a necessary rate. This intuitive lower
       bound can be improved by examining the probability of correctly
       inferring the random path $W^t$ from observing the vector process $\XX^t$.
       This probability is related to the following event:
       \begin{equation}\label{eq:alternating_prob}
        A_L :=\{ \ (X_0,\ldots, X_{L}) = (x_0, x_1 , x_0, x_1,
       \ldots),\  x_0, x_1 \ \in {\cal X}   \}.
       \end{equation}
       The probability of the event $A_L$ is closely related to
       the probability of ambiguity from the observation, making
       the trajectory unidentifiable. To see this,
       let $L=4$ and consider inferring $W_1$ from the observation of $(\XX_0, \XX_1)$. If
       $\XX_0 = (x_0,x_1,x_0,x_1)$ and  $\XX_1 = (x_1,x_0,x_1,x_0)$, then it follows that $W_1$
       cannot be unambiguously determined from $(\XX_0,\XX_1)$. Intuitively, if $W^t$ can be
         determined from $\XX^t$, then the complexity of the
       trajectory is embedded in $\XX^t$ and thus independently adds to the information complexity of $X$. If, however, there is ambiguity on
        $W^t$, then sets of $W^t$ that are consistent with a particular trajectory $\mathbf{v}^t$ can be indexed and
        coded with a lower rate. We are now ready to state and prove Theorem 1.

        \begin{theorem} \label{THM:LB_UB_V}
        Consider the vector process $\XX$ consisting of $L$-tuples
        generated by a Bernoulli random walk with transition probability $p_W \leq 1/2$, and a
        wall process $X$ drawing values i.i.d. on a finite alphabet, and that has entropy $H(X)$.
        The conditional entropy $H(\XX_t|\XX^{t-1})$ obeys
        \begin{equation}\label{eq:LB_UB_conditional}
          \prob{\overline{\recset^t}}H(X) +   H(p_W) - H(P_e)  \leq   H(\XX_t|\XX^{t-1})   \leq
        \frac{1}{t}\sum_{i=1}^t \prob{\overline{\recset^i}}H(X) +
        H(p_W),
        \end{equation}
        where $P_e$ is the probability of error in estimating $W_1$ from observing $(\XX_1,\XX_0)$. The entropy rate $H(\XX)$ satisfies
       \begin{equation} \label{eq:LB_UB}
       (1-2p_W)H(X) + H(p_W) - H(P_e) \leq  H(\XX)  \leq
        (1-2p_W)H(X) + H(p_W).\\
       \end{equation}%
       \end{theorem}


       \begin{proof}
       For each $t$ we have
       \begin{eqnarray} \label{eq:proof_theorem1_eq1}
       H(\XX_t|\XX^{t-1}) & \stackrel{\small (a)}{\leq} &  \frac{1}{t} \sum_{i=1}^tH(\XX_i|\XX^{i-1})=\frac{H(\XX^t)}{t} \nonumber \\
                          & \stackrel{\small (b)}{\leq} &   \frac{H(\XX^t)  +  H(W^t|\XX^t)}{t} \label{eq:4.7} \\
                          & =&  \frac{H(W^t)  +  H(\XX^t|W^t)}{t}  \\
                          & =&  \frac{H(W^t) +  \sum_{i=1}^tH(\XX_i|\XX^{i-1},W^t)}{t} \nonumber   \\
                          & \stackrel{\small (c)}{=}     &  H(p_W) + \frac{1}{t} \sum_{i=1}^t H(\XX_i|\XX^{i-1},W^i) \label{eq:4.9},
       \end{eqnarray}
       where (a) follows because  $H(\XX_t|\XX^{t-1})$ decreases with $t$, (b) holds because $H(W^t|\XX^t) \geq 0$, and (c) is true because $H(W^t)=tH(p_W)$ and $(W_{i+1},\ldots,W_t)$ is independent of $(V^i,W^i)$.
       Further, it is true that
       \begin{eqnarray*}
       H(\XX_i|\XX^{i-1},W^i=w^i, w^i  \ \text{is recurrent} ) &=& 0.\\
       H(\XX_i|\XX^{i-1},W^i=w^i, w^i  \ \text{is not recurrent} ) &=& H(X).
       \end{eqnarray*}
       Consequently,
       \begin{eqnarray} \label{eq:condi_entropy2}
       H(\XX_i|\XX^{i-1},W^i)& =&  \sum_{w^i\in\overline{\recset^i}} \prob{W^i=w^i}H(\XX_i|\XX^{i-1},W^i=w^i) \nonumber \\
       &=&  \prob { \ \overline{\recset^i}} H(X).
       \end{eqnarray}

        Combining (\ref{eq:proof_theorem1_eq1}) and
       (\ref{eq:condi_entropy2}) gives the upper bound in (\ref{eq:LB_UB_conditional}). We now turn to the lower bound. Using the chain rule for mutual
       information and the information inequality, we have
       \begin{eqnarray}\label{eq:lower_bound_discrete1}
       H(\XX_t|\XX^{t-1}) & = & H(\XX_t|\XX^{t-1},W^t) + I(W^t;\XX_t|\XX^{t-1}) \nonumber \\
                          & = & H(\XX_t|\XX^{t-1},W^t) + I(W^{t-1};\XX_t|\XX^{t-1}) + I(W_t;\XX_t|\XX^{t-1},W^{t-1}) \nonumber \\
                          & \geq & H(\XX_t|\XX^{t-1},W^t) + I(W_t;\XX_t|\XX^{t-1},W^{t-1}).
       \end{eqnarray}
       Moreover, because the random walk increment $W_t-W_{t-1}$ is independent
       of $(\XX^{t-1},W^{t-1})$, it follows that
       \begin{eqnarray}\label{eq:lower_bound_discrete2}
       I(W_t;\XX_t|\XX^{t-1},W^{t-1}) & = & H(W_t|\XX^{t-1}, W^{t-1}) - H(W_t|\XX^{t},W^{t-1}) \nonumber \\
                                      & = & H(p_W) - H(W_t|\XX^{t},W^{t-1}).
       \end{eqnarray}
       We proceed by finding an upper bound for
       $H(W_t|\XX^{t},W^{t-1})$. Because conditioning reduces entropy, and using Markovianity, we have that
       \begin{eqnarray}
       H(W_t|\XX^{t},W^{t-1}) &\leq & H(W_t|\XX_{t},\XX_{t-1},W_{t-1} ) \nonumber \\
        &=&  H(W_1|\XX_{1},\XX_{0}).
       \end{eqnarray}%
       Denote by $P_e$ the probability of error of estimating $W_1$ from observing $(\XX_1, \XX_0)$. Then, Fano's inequality
        gives that
       \begin{equation}
       H(W_1| \XX_{1},\XX_{0} ) \leq  H(P_e) + \log_2(1).
       \end{equation}
       Combining this with
       (\ref{eq:lower_bound_discrete1} - \ref{eq:lower_bound_discrete2})
       and (\ref{eq:condi_entropy2}), we assert
       the lower bound  in   (\ref{eq:LB_UB_conditional}). By letting $t\rightarrow \infty$ in   (\ref{eq:LB_UB_conditional}) and using Lemma \ref{LEMMA:RWALK}
       (i) we obtain (\ref{eq:LB_UB}).
      \end{proof}



        %

        \begin{remark}
        The upper bound of Theorem \ref{THM:LB_UB_V} contains slack. One trivial
        example is when the entropy of the process ${X}$ is $0$. In such
        a case the upper bound reduces to $H(p_W)$, which is clearly loose given
        that the vector process $\XX$ has zero entropy in this case.
        The size of the conditional entropy
        $H(W_t|\XX^t)$ determines the amount of slack in the
        bounds (see (\ref{eq:4.7})). Such entropy depends, among other things, on the size of the alphabet of the process
        $\XX$ and on the block length $L$, as the next example illustrates.
        \end{remark}

        \begin{remark}
        In the case where $L$ is odd, then an expression for the slack in terms of the probability of the
        set $A_L$ in (\ref{eq:alternating_prob}) can be obtained.  Denote by ${\cal A}_1$
        the set of $(\mathbf{v}_1, \mathbf{v}_0)$ is such that $W_1$ cannot be
        inferred with with probability one. Then, because $L$ is odd,
        it is straightforward to infer that
        \footnote{Note that when $L$ is even, then we cannot assert that $\prob{W_1=1|(\mathbf{v}_0,\mathbf{v}_1) \in {\cal A}_1 }  = p_W$. }
       $\prob{W_1=1|(\mathbf{v}_0,\mathbf{v}_1) \in {\cal A}_1 }  = p_W$ and
       $\prob{W_1=1|(\mathbf{v}_0,\mathbf{v}_1) \in \overline{{\cal A}_1 }} = 0$.
       Consequently, we have that
        \begin{eqnarray}
        H(W_1|\XX_0, \XX_1)& = & \prob{ {\cal A}_1 } H\left ( \prob{W_1=1|(\mathbf{v}_0,\mathbf{v}_1) \in {\cal A}_1 }\right ) \\ \nonumber
        & = & \prob{ {\cal A}_1 }  H(p_W).
       \end{eqnarray}
       The  set  ${\cal A}_1$ is
       contained in the set $\{ \XX_{t-1} = (x_0,x_1,\ldots), \XX_{t} =
       (x_1,x_0,\ldots)\}$. Therefore, we have that $\prob{{\cal A}_1} \leq \prob{ A_L}$ and so $H(W_1|\XX_0 , \XX_1) \leq \prob{ A_L}H( p_W)$.
       Combining this with (\ref{eq:lower_bound_discrete1} - \ref{eq:lower_bound_discrete2})
       and (\ref{eq:condi_entropy2}), we obtain the following bound:
       \begin{equation} \label{eq:LB_UB_AL}
       (1-2p_W)H(X) + H(p_W)\prob{\overline{A_L}} \leq  H(\XX)  \leq
       (1-2p_W)H(X) + H(p_W).\\
       \end{equation}
        This special case of the results in Theorem 1 is useful because the slack can be computed analytically as in Example 1 below.
%
        \end{remark}
%

        \begin{remark}
        Note that for any $L$, we always have that $P_e \leq \prob{ A_L}$ so that in many cases, as $L\rightarrow \infty$,
        then  $\prob{ A_L}\rightarrow 0$, and the bounds in Theorem 1 become tight. Theorem \ref{THM:LB_UB_V} shows that, under some
        conditions, optimal encoding in the information-theoretic
        sense can be attained by extracting and optimally coding the trajectory
        $W^t$,  and optimally coding the spatial innovations in the vector samples $\XX^t$. 
        \end{remark}
        
        \begin{remark}
        For the symmetric random walk case, there is an intuitive explanation for the $H(p_W)$ upper bound. At time $t$, with high probability we have that $-c \sqrt{t} < W_s < c \sqrt{t}$ for $0 \leq s \leq t$. Therefore, with high probability, the number of new sites that are visited up to time $t$, which is $\frac{2}{t} \max _{0\leq s\leq t} |W_s|$, is less than $\frac{2}{\sqrt{t}}$ which converges to zero as $t\rightarrow \infty$. Thus, the term corresponding to the entropy rate of the source $X$ vanishes in the information rate of the $V$ process. \\
        \end{remark}

\begin{figure}[h!]\centering
         {\includegraphics[width = 11cm]{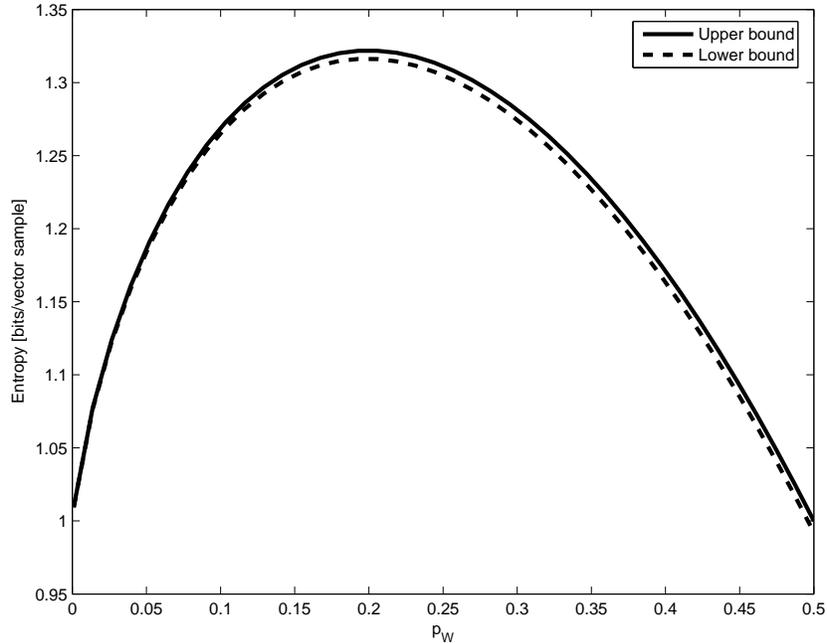}}
        \caption{Bounds on information rate.  Lower and upper bounds as a function of $p_W$
        for the binary wall with $p_X=1/2$ and $L=9$. \label{Fig:bounds_them1}}
        \end{figure}

        \begin{example}Suppose that the $X$ is uniformly distributed
        over $|{\cal X}|$ values. Then, it
        is easily seen that $$\prob{A_L} = \frac{1}{  |{\cal X}|^{L-1} }.$$
        Consequently, the difference between upper and lower bounds in (\ref{eq:LB_UB_conditional}) decays
        exponentially fast when the block length $L \rightarrow \infty$. For
        a fixed $L$, the difference also decays as $|{\cal X}|$
        increases. Thus, for  $L$ and $|{\cal X}|$ sufficient large, we have that $\prob{A_L}\approx 0$, and we can
        approximate the entropy rate as $$H(\XX) \approx (1-2p_W) \log|{\cal X}| + H(p_W)$$ bits
        per block. Note that if $p_W=1/2$, then the recurrence property of the random walk generates redundancy that has the effect of reducing the entropy of the vector process. Figure  \ref{Fig:bounds_them1} illustrates the bounds when $X$ is Bern(1/2), and
        $L=9$. We see that in this case, the derived upper and lower bounds are very tight. 
        \end{example}

        \subsection{Memory constrained coding }

        From source-coding theory, the entropy rate $H(\XX)$ can be attained with an encoder-decoder pair with
        unbounded memory and delay. In the finite memory case,
         often the encoder has to code $\XX_t$ based on
        the observation of $\XX_{t-1},\dots, \XX_{t-M}$, and the decoder proceeds accordingly. This
        situation is similar to one encountered in video compression, where
        a frame at time $t$ is coded based on $M$ previously coded frames \cite{Telkap:95bk}. In this
        case, the  average code-length is  bounded below by the
        conditional entropy $H(\XX_{t}|\XX_{t-1}, \ldots, \XX_{t-M}) =
        H(\XX_M|\XX_{M-1},\ldots,\XX_0)$.
        The bound (\ref{eq:LB_UB_conditional}) in Theorem \ref{THM:LB_UB_V}  describes the
        behavior of the conditional entropy $H(\XX_M|\XX^{M-1})$. Intuitively,
        by looking at the stored samples from $t-M$ up to
        $t$, the encoder can separately code $W_t$ and take advantage of
        recurrences present from $t-M$ to $t-1$. In effect, finite memory
        prevents the encoder to exploit long term recurrences that are not
        visible in the memory. Similar observations are verified in practice for
        instance in \cite{Girod:99,ForchheimerLi:96,CHerley:06}.

\begin{figure}[h!]\centering
        \center
        \subfigure[]{\includegraphics[width = 11cm]{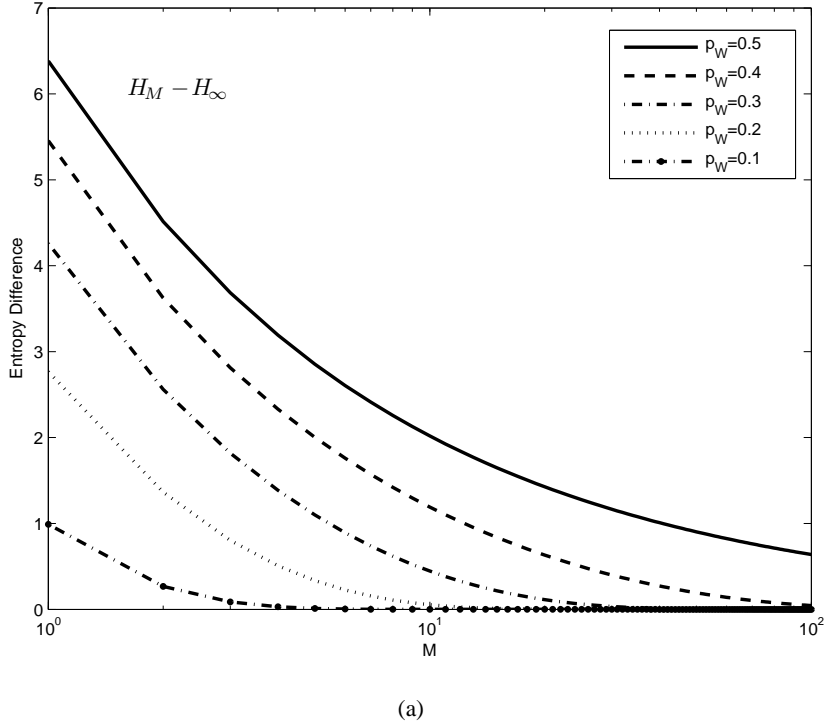}}
        \caption{Memory constrained coding. Difference $H(\XX)-H(V_M|V^M)$
         as a function of $M$. When $p_W=0.5$, the bit rate can be lowered significantly at the cost of a large memory. A moderate
         bit rate reduction is obtained with small values of $M$ when $p_W=0.1$. The curves are computed
         using Theorem \ref{THM:LB_UB_V} for $X$ uniform over an alphabet of size 256. \label{Fig:memory}}
        \end{figure}

        Figure  \ref{Fig:memory} illustrates
        how memory influences coding when $X$ is uniform over
        an alphabet of size $|{\cal X}|=256$.
        The curves are computed using the upper bound in (\ref{eq:LB_UB_conditional}). Because the alphabet
        size is large, the bound is tight.
        In the most recurrent case with
        $p_W=0.5$, the conditional entropy approaches the
        entropy rate at a slower rate when $M\rightarrow \infty$ [see (\ref{eq:LB_UB_conditional}-\ref{eq:LB_UB})]. Furthermore,
        as $M$ approaches infinity, there is a significant reduction in the
        conditional entropy. For instance, an encoder that uses 1 frame in
        the past with optimal coding would need about twice as many bits
        as one that uses 4 frames. By contrast, when $p_W=0.1$, because
        longer term recurrences are rare, moderate values of $M$ are already
        enough to attain the limiting rate. As a result there is little to
        gain by increasing $M$.

        The observations drawn from Figure  \ref{Fig:memory} are also verified in practice for
        instance in \cite{ForchheimerLi:96,Vasconcelos:94,Girod:99}. Finally, we point out that the issue of exploiting
        long term recurrences dates back to Ziv-Lempel \cite{ZivLempel:78}
        in lossless compression. Extension of the Lempel-Ziv algorithm to
        the lossy case is discussed in \cite{Gutman:93}, and
        lossless compression of two-dimensional array in \cite{ZivLempel86}. More
        recently, an universal scheme to optimally scan and predict data in
        a multidimensional field with applications to video is presented in
        \cite{CohenWeissman:07}.

         \subsection{Lossy information rates}

         In this section we assume again that the process $X$ is i.i.d. and that $X_n$ takes values over a finite alphabet ${\cal
         X}$.
         Information rates for the lossy case take the form of a
         rate-distortion function. Consider a
         $t$-tuple $(\XX_1, \ldots ,\XX_t)$ where each $\XX_j$ is a
         random vector taking values in ${\cal X}^L$. A reproduction $t$-tuple is denoted by $(\hat \XX_1, \ldots, \hat
         \XX_t)$, and its entries take values on a reproduction alphabet $\hat {\cal X}$. A distortion measure is defined as follows:
         $$d(\XX^t, \hat \XX^t) = \frac{1}{tL} \sum_{i=1}^t d_s(V_i,\hat V_i),$$
         where $d_s:{\cal X}\times{ \cal \hat X}\rightarrow \mathbb{R}^+$ is a distortion measure for an
         $L$-dimensional vector. For example, for the MSE metric we have $d(V_i,\hat V_i) = \|V_i - \hat       V_i\|^2$.

         The rate-distortion function for each $t$, and for given
         distortion measure, is written as
        \begin{equation}\label{eq:rate-distortion}
        R_{\XX^t}(D) = \inf_{\mathbb{E}d(\XX^t,\hat \XX^t)\leq D}
        \frac{ I(\XX^t;\hat{\XX}^t)}{t},
        \end{equation}
        where the infimum of the normalized mutual information $\frac{
        I(\XX^t;\hat{\XX}^t)}{t}$ is taken over all conditional pdf's $\prob{\hat V^t|V^t}$  such that $\mathbb{E}d(\XX^t,\hat \XX^t)\leq D$.

        The rate-distortion function for the
        process $\XX = (\XX_1, \XX_2, \ldots)$ is given by \cite{Berger:71}
        \begin{eqnarray}
        R_{\XX}(D) = \lim_{t\rightarrow \infty} R_{\XX^t}(D).
        \end{eqnarray}
        Because the process $\XX$ is stationary, it can be shown that the above limit  always
        exists (see \cite[p. 270]{Berger:71}, or \cite{Pinsker84bk}).


        By coding the side information $W^t$ separately,
        an upper bound for $R_{\XX}(D)$ similar to Theorem 1 can be
        developed. The upper bound is based on the notion of
        {\it conditional rate-distortion}
        \cite{Gray:73,Berger:71}. This notion is developed in the lemma below.

        \begin{lemma}  \label{LEMMA:conditionalRD} (Gray \cite{Gray:73})
        Let $\XX$ be a random vector taking values in ${\cal X}$ and let
        $W$ be another random  variable. Define the conditional
        rate-distortion:
       \begin{equation}
       R_{\XX|W}(D) = \inf_{\mathbb{E}d(\XX,\hat \XX)\leq D}
     I(\XX;\hat{\XX} |W  ),
       \end{equation}
       \noindent where the infimum is taken over all conditional
       distributions of $\hat \XX$ given $ \XX$ and $W$. The conditional rate-distortion obeys
       \begin{equation}
        R_{\XX|W}(D)\leq R_{\XX}(D) \leq R_{\XX|W}(D) +
        I(\XX;W).\\
        \end{equation}
        \end{lemma}

        The conditional rate-distortion
        of $\XX^t$ conditioned on $W^t$ is defined as follows:

        \begin{equation}\label{eq:conditional_RD_def}
        R_{V^t|W^t}(D) = \inf_{\mathbb{E}d(\XX^t,\hat \XX^t)\leq D}
        \frac{ I(\XX^t;\hat{\XX}^t|W^t)}{t},
        \end{equation}
        \noindent where the infimum is taken over all probability distributions of $\hat \XX^t$ conditional on $ \XX^t$
        and  $W^t$. The conditional rate-distortion can be bounded in
        terms of the rate-distortion function of the process $X$.

        \begin{proposition} \label{PROP:conditional_RD_V}
        The conditional rate-distortion function satisfies
        \begin{equation} \limsup_{t\rightarrow \infty} R_{\XX^t|W^t}(D)\leq (1-2 p_W)R_X(D).\end{equation}
        \end{proposition}
        \begin{proof}
        Let $\lambda(w^t)$ denote the number of new sites from the path
        $w^t$. Then, conditional on $w^t$, the $\XX^t$ has only
        $\lambda(w^t)$ entries that need to be encoded. For each
        $(w^t,v^t)$, let $f_{w^t}(v^t)$ denote the vector with the
        $\lambda(w^t)$ entries of $v^t$ to be coded. Moreover, let $\XX$
        and $\hat \XX$ be such that $\mathbb{E}d_s(\XX_i[j] ,\hat \XX_i[j]) \leq D$ for $i=0,\ldots,t$, and $j=0,\ldots,L-1$. We have
        \begin{eqnarray}
        I(\XX^t;\hat \XX^t |W^t)  &=& \sum_{w^t} \prob{W^t=w^t} I(\XX^t;\hat \XX^t |W^t=w^t) \nonumber \\
                       &\geq& \sum_{w^t} \prob{W^t=w^t} I(f_{w^t}(\XX^t); f_{w^t}(\hat \XX^t) |W^t=w^t)\nonumber  \\
                        &\geq& \sum_{w^t} \prob{W^t=w^t} \lambda(w^t) R_X(D) \nonumber \\
                        & = & \mathbb{E}\lambda(W^t) R_X(D) \label{eq:cond_RD_B},
        \end{eqnarray}
        where we have used the inequality $I(X;Y) \geq I(f(X);g(Y))$ for
        measurable functions $f,g$ \cite{Gray07bk}, and the fact that the
        process $X$ is i.i.d. and independent of $W^t$, and that the
        individual distortions are less than $D$. The lower bound can be
        achieved as follows. Let $p_*(\hat X|X)$ be the test channel that
        attains $R_X(D)$. We let  $\hat X_n$ be the result of passing
        $X_n$ though the channel $p_*(\hat X_n|X_n)$. For each given $w^t$
        we construct $\hat V^t$ from $\hat X$ and $w_t$. This results in a
        joint conditional distribution that attains the lower bound
        (\ref{eq:cond_RD_B}).

        Because the lower bound is attainable, it follows that
        $$R_{\XX^{t}|W^t}(D)\leq \frac{\mathbb{E}\lambda(W^t)}{t} R_X(D).$$
        Moreover, using Lemma \ref{LEMMA:RWALK} it is straightforward to
        check that $t^{-1} \mathbb{E}\lambda(W^t)$ converges to
        $(1-2p_W)$, which concludes the proof.
        \end{proof}

        The above proposition enables us to derive an upper bound
        for the rate-distortion function.

        \begin{theorem} \label{THM:LOSSYSTATIC}Consider the i.i.d. process $X$ such
        that $X_n$ takes values over a finite alphabet ${\cal X}$.
        Let $R_X(D)$ denote its rate-distortion function.
        The rate-distortion function of the process $V$ satisfies
        \begin{equation}\label{eq:BOUNDS_CONT_AMPLI} 
        R_V(D) \leq H(p_W) + (1-2p_W)R_X(D). \end{equation}
        \end{theorem}
        \begin{proof}
        Using Lemma \ref{LEMMA:conditionalRD} we have the
        following bound based on the  conditional rate-distortion function
        \cite{Gray:73}:
        \begin{eqnarray*} R_{\XX^t|W^t}(D)\ \leq \  R_{\XX^t}(D) &\leq & R_{\XX^t|W^t}(D) + \frac{1}{t}I(\XX^t;
        W^t)\\ &  \leq & R_{\XX^t|W^t}(D) + H(p_W).
        \end{eqnarray*}
        Letting $t\rightarrow \infty$ and using Proposition \ref{PROP:conditional_RD_V} asserts (\ref{eq:BOUNDS_CONT_AMPLI}).
        \end{proof}
        
        \begin{remark}
        To describe $V^t$ to the decoder with average expected distortion less than $D$ we do as follows. Covey the trajectory $W^t$ to the decoder spending on average $\approx t H(p_W)$     bits.  Then describe the ``spatial innovations'' with an average expected distortion less than $D$ spending $\approx \lambda_t R_{\cal X}(D)$ bits where $\lambda_t \approx t  (1-2p_W)$ is the number of new sites visited up to time $t$. On average, by using this scheme, one needs $H(p_W)+(1-2p_W)R_{\cal X}(D)$ bits which is the upper bound presented in (\ref{eq:BOUNDS_CONT_AMPLI}).
        \end{remark}
        
        \begin{remark}
        Because the alphabet is finite, if the reproduction alphabet $\hat{\cal  X}$ is a superset of the original
        alphabet ${\cal X}$ and, in addition, the distortion measure is such that $d(x, \hat x)=0$ if and only if $x=\hat x$, then  we have that for each
        $t$, $R_{\XX^t}(D)$ converges to $ {t^{-1}}{H(\XX^t)}$
        as $D \rightarrow 0$ \cite{Berger:71}. Consequently, for large alphabet
        sizes and large block length, the entropy rate bound of (\ref{eq:BOUNDS_CONT_AMPLI})
        is sharp, and so the above bound on the rate-distortion is also sharp for small
        distortion values.
         \end{remark}

%

        Theorem \ref{THM:LOSSYSTATIC} shows that in the low
        distortion regime, optimal encoding in the information-theoretic sense can be
        attained by extracting and coding $W^t$ losslessly, and using the remaining bits to optimally code
        the vector samples corresponding to spatial innovations.
        This statement has implications, for example for high rate
        video coding, since it indicates that motion should be
        encoded exactly, with the remaining bits allocated to
        prediction errors.


                \section{Information Rates for Dynamic Reality}  \label{SecIV:innovation}
                \subsection{Model}
                The model in the previous section assumes a ``static background.'' More precisely, the infinite wall process $X$ is drawn at time $0$ and does not change after that.  In practice, however,
                scene background changes with time and a suitable model would have to account
                for those changes. New information comes fundamentally in two  forms: the first
                consists of information that is ``seen'' by the camera for the first time, while the second consists of changes
                to old information (e.g., changes in the background). In this section, we propose a model that accounts for both
                 these sources of new information.

                To develop a model for scenes that change over time, we model $X$ as
                a 2-D  random field indexed by $(n,t)\in\mathbb{Z}\times\mathbb{Z}^+$.
                A simple yet rich model for the field is that of a first order Markov model over time, and i.i.d. in
                space. The random field is defined as follows:

                \begin{definition} The {\it random field} is the field $RF = \{X^{(t)}_n : (n,t)\in\mathbb{Z}\times\mathbb{Z}^+    \},$
                such that $(X^{(0)}_n:n\in\mathbb{Z})$ is i.i.d. and for each $n \in \mathbb{Z}$, the process  $(X^{(t)}_n:t\in\mathbb{Z})$ is a first
                order time-homogeneous Markov process possessing a stationary distribution.
                \end{definition}

                 The fact that the random field  $(X^{(0)}_n:n\in\mathbb{Z})$ is i.i.d. simplifies
                 calculations considerably. One justification for this model is when
                the field is Gaussian. In such case, independence is attained by a simple linear
                transformation of the process $(X^{(0)}_n:n\in\mathbb{Z})$. It can be shown that
                such transformation preserves the Markovianity on the time dimension, and the i.i.d.
                assumption can be justified in this case.

                Throughout this section, we assume that the Markov chain of the vector
                process is already in steady-state. This assumption is common, for example,
                in calculating rate-distortion functions for Gaussian processes with memory \cite{Berger:71}.

                The {\it dynamic vector process} $\XX$ is defined similarly to the static case,
                but now taking snapshots or vectors from the random field:
                \begin{definition}
                Let $RF = \{X^{(t)}_n : (n,t)\in\mathbb{Z}\times\mathbb{Z}^+    \}$ be a random field, and let $W$ be a random walk.
                The dynamic vector process is the process $\XX=(\XX_t:t\in\mathbb{Z}^+)$ such that for each $t>0$,
                $$\XX_t = (X_{W_t}^{(t)},X_{W_t+1}^{(t)},\ldots,X_{W_t+L-1}^{(t)}).$$
                \end{definition}

\begin{figure}[htb]
                 \centering
                 \psfrag{t}{$t$}
                 \psfrag{n}{$n$}
                 \psfrag{ldots}{$\cdots$}
                 \psfrag{vdots}{\hspace{0.3cm}$\vdots$}
                 \psfrag{X10}{\hspace{-.15cm} \tiny $X_{n-1}^{(t)}$}
                 \psfrag{X11}{\hspace{-.15cm} \tiny $X_{n}^{(t)}$}
                 \psfrag{X12}{\hspace{-.15cm} \tiny $X_{n+1}^{(t)}$}
                 \psfrag{X20}{\hspace{-.15cm} \tiny $X_{n-1}^{(t+1)}$}
                 \psfrag{X21}{\hspace{-.15cm} \tiny $X_{n}^{(t+1)}$}
                 \psfrag{X22}{\hspace{-.15cm} \tiny $X_{n+1}^{(t+1)}$}
                 \psfrag{X30}{\hspace{-.15cm} \tiny $X_{n-1}^{(t+1)}$}
                 \psfrag{X31}{\hspace{-.15cm} \tiny $X_{n}^{(t+2)}$}
                 \psfrag{X32}{\hspace{-.15cm} \tiny $X_{n+1}^{(t+2)}$}
                 \subfigure[]{\includegraphics[width=6.5cm ]{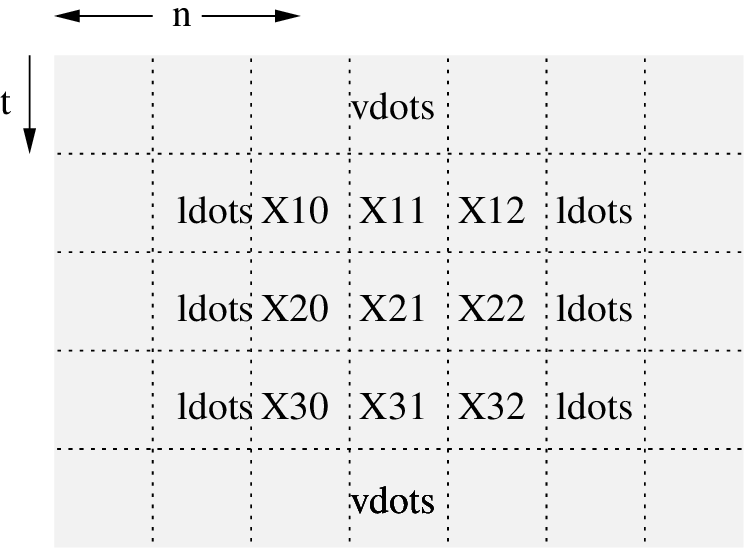}} \ \ \ \
                 \psfrag{t}{$t$}
                 \psfrag{n}{$n$}
                 \psfrag{ldots}{ }
                 \psfrag{vdots}{\hspace{.2cm} $\vdots$}
                 \psfrag{X00}{\tiny $X_{0}^{(0)}$}
                 \psfrag{X01}{\tiny $X_{1}^{(0)}$}
                 \psfrag{X02}{\tiny $X_{2}^{(0)}$}
                 \psfrag{X10}{\tiny $X_{1}^{(1)}$}
                 \psfrag{X11}{\tiny $X_{2}^{(1)}$}
                 \psfrag{X12}{\tiny $X_{3}^{(1)}$}
                 \psfrag{X20}{\tiny $X_{0}^{(2)}$}
                 \psfrag{X21}{\tiny $X_{1}^{(2)}$}
                 \psfrag{X22}{\tiny $X_{2}^{(2)}$}
                 \psfrag{X30}{\tiny $X_{0}^{(3)}$}
                 \psfrag{X31}{\tiny $X_{1}^{(3)}$}
                 \psfrag{X32}{\tiny $X_{2}^{(3)}$}
                 \subfigure[]{\includegraphics[width=6.5cm ]{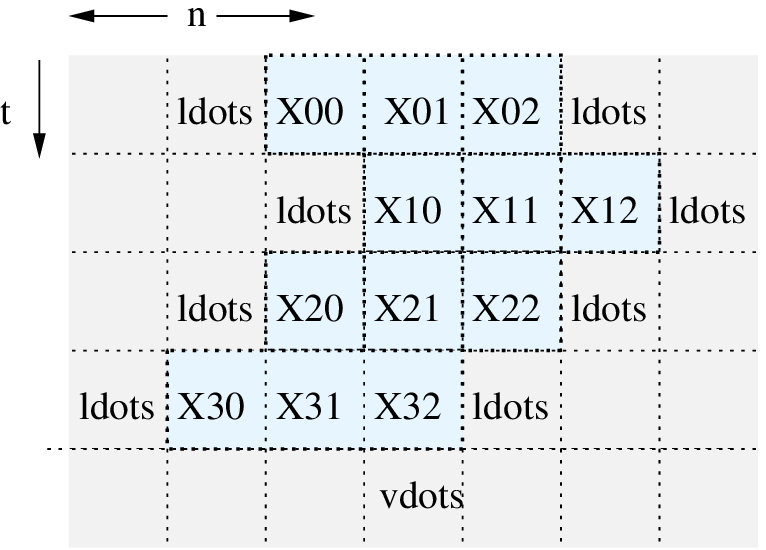}}
                 \caption{A model for the dynamic reality. (a) First, there is a random field  that is Markov in the time dimension $t$,
                 and i.i.d. in the spatial dimension $n$. (b) Motion then occurs within this random field. \label{Fig:Innovation}}
                 \end{figure}

                The random field and the corresponding vector process are
                illustrated in Figure  \ref{Fig:Innovation}.
                 
                 We point out here that the proposed random field model is a simplified depiction of real visual scene data. For instance, we acknowledge that the spatial independence assumption in non-Gaussian cases is not met in practice, and that the camera motion is not i.i.d. in practice. We stress however that true rate-distortion bounds are difficult to derive for more elaborated sources, and that even a simplified model with true coding bounds is useful provided its deficiencies are acknowledged.  
                 

                 \subsection{Lossless information rates }

                 In the development that follows we assume, for simplicity, that
                 the random field takes values on a finite
                 alphabet ${\cal X}$. The results can equally be developed for a
                 random field taking values over $\mathbb{R}$, under suitable
                 technical conditions.

                 To derive bounds for $H(\XX)$ in the dynamic reality case, we compute the
                 following conditional entropy rate:
                 \begin{equation} \label{eq:conditional_entropy_def}
                 H(\XX|W) :=  \lim_{t\rightarrow
                 \infty}H(\XX_t|\XX^{t-1},W^{t}),\end{equation}
                 if {\it the limit exists}. As we shall see
                 in the examples that follow, the above limit can be computed analytically.
                 The key is to compute $H(V_t|V^{t-1},W^t=w^t)$ by
                 splitting the set of all paths into recurrent and nonrecurrent paths, and
                 further splitting the set of recurrent paths
                 according to (\ref{eq:disjoint_union}).

                 Referring to Figure \ref{Fig:Innovation}(b), let $w^t$ be a given path and consider the process
                 $\XX^t$. Note that each $\XX_t$
                 has $L-1$ entries from the same spatial location as $L-1$ entries from $V_{t-1}$. The
                 remaining entry corresponds to either a nonrecurrent or
                 a recurrent location depending on $w^t$. If $w^t$ is nonrecurrent,
                 then by the Markov property of the
                 field, we have  $$H(V_t|V^{t-1}, W^t=w^t) = H(X_0^{(t)}) + (L-1)H(X_0^{(t)}|X_0^{(t-1)}).$$ If a path
                 is recurrent at $t$, then there is an $s<t$ such that $w_s = w_t$ but $w_{t} \neq w_i,$ for $s<i<t$.
                 Using the Markov property again,
                 it follows that $H(V_t|V^{t-1}, W^t=w^t) =H(X_0^{(t)}|X_0^{(s)}) + (L-1)H(X_0^{(t)}|X_0^{(t-1)})$.
                 The above argument is explicitly written as follows:
                 \begin{eqnarray}\label{eq:conditional_entropy_innovations}
                 H(\XX_t|\XX^{t-1},W^t) & = & \sum_{w^t \in \overline{\recset^t}}
                 H(\XX_t| \XX^{t-1},W^t=w^t) \prob{W^t=w^t} \nonumber \\ &\ \ \ &\ \ \ \ \ \ \ + \sum_{ w^t \in \recset^t} H(\XX_t|\XX^{t-1}, W^t=w^t )\prob{W^t=w^t}  \\
                 & = & \left( H(X_0^{(t)}) + (L-1)H(X_0^{(t)}|X_0^{(t-1)})  \right )
                  \prob{\overline{\recset}^t} \nonumber \\
                  &\ \ \  & \ \ \ \ \ \  + \sum_{i=1}^{\lfloor t/2 \rfloor}
                                \sum_{w^t \in {\cal T}_{t-2 i}^{t}} H(\XX_t|\XX^{t-1},W^t=w^t) \prob{W^t=w^t}  \\
                 & = & \left( H(X_0^{(t)}) + (L-1)H(X_0^{(t)}|X_0^{(t-1)})    \right)
                 \prob{\overline{\recset}^t} \nonumber \\
                 & \ \ \ & \ \ \ \ \ \  + \sum_{i=1}^{\lfloor t/2 \rfloor}
                 \left[\ (H(X_0^{(t)}|X_0^{(t-2i)}) + (L-1)H(X_0^{(t)}|X_0^{(t-1)}) \right )  \prob{{\cal T}_{t-2i}^{t}}   \\
                 & = & (L-1)H(X_0^{(t)}|X_0^{(t-1)}) + H(X_0^{(t)})
                 \prob{\overline{\recset}^t} \nonumber \\
                 & \ \ \ & \ \ \ \ \ \ \ + \sum_{i=1}^{\lfloor t/2 \rfloor}
                 H(X_0^{(2i)}|X_0^{(0)}) \prob{{\cal T}_0^{2i}}. 
                 \end{eqnarray}
                 By letting $t \rightarrow \infty$ using Lemma \ref{LEMMA:RWALK} (i) leads to
                 \begin{equation}\label{eq:condentropy_rate_innv}
                 H(\XX|W) =   H(X_0^{(\infty)}) (1-2p_W) + (L-1)H(X^{(1)}_0|X^{(0)}_0) + \sum_{i=1}^{ \infty } H(X^{(i)}_0|X^{(0)}_0)
                 \prob{\retset^{i}},
                 \end{equation}
                 where $\prob{\retset^{ i}}$ is the probability of return given in Lemma \ref{LEMMA:RWALK} (ii).               
                 The infinite sum in the left-hand side of
                 (\ref{eq:condentropy_rate_innv}) is well-defined. It is an infinite sum of positive numbers, and it is bounded above by
                 $H(X_0^{(\infty)}) \sum_{i=1}^{ \infty } \prob{\retset^{i}} = H(X^{(\infty)}) 2p_W.$ Note that we replaced $2 i$ with $i$ in the infinite sum above in view of the fact that $\prob{{\cal T}^{2i+1}}=0 $.

                 With the conditional
                 entropy rate in (\ref{eq:condentropy_rate_innv}) we can derive lower and upper bounds on the
                 entropy rate $H(V)$. To derive an upper bound,
                 we bound $\frac{H(V^t)}{t}$ for each $t$ and
                 let $t\rightarrow \infty$. For the lower bound, similar to Section \ref{SecIII:lossless}, we bound $H(\XX_t|\XX^{t-1})$ below.
                 Because the alphabet ${\cal X}$ is finite
                 and the process is stationary, the limits of $\frac{H(V^t)}{t}$ and $H(\XX_t|\XX^{t-1})$
                 as $t\rightarrow \infty$ coincide.

                 The upper bound is obtained from the
                 inequality $H(\XX^t) \leq t H(p_W)+H(\XX^t|W^t)$.
                 Note that $H(\XX^t|W^t)=\sum_{i=1}^t H(V_i|\XX^{i-1},W^t)$, so that if $H(V_i|\XX^{i-1},W^t)$
                  converges to a limit as $t\rightarrow \infty$, we have necessarily
                  that $t^{-1} H(\XX^t|W^t)$ converges to the same limit (see
                  e.g., \cite[p. 64]{CoverThomas:91}). So,
                  \begin{eqnarray*}
                  \lim_{t\rightarrow \infty}
                  \frac{H(\XX^t)}{t} &\leq& H(p_W) + \lim_{t\rightarrow
                  \infty}\frac{H(\XX^t | W^t )}{t}\\
                  &=& H(p_W) +\underbrace{ \lim_{t\rightarrow
                  \infty} H(\XX_t | \XX^{t-1}, W^t )}_{H(V|W)}.
                  \end{eqnarray*}

                  To derive a lower bound, note that the
                  development leading to
                  (\ref{eq:lower_bound_discrete1}-\ref{eq:lower_bound_discrete2}) for the static case also holds for the dynamic
                  case. So, we have
                  \begin{equation} \label{eq:dynamic_lowerB} H(\XX^t|\XX^{t-1}) \geq H(p_W) + H(\XX_t|\XX^{t-1},W^t) -
                  H(W_t|\XX^t,W^{t-1}).\end{equation}
                 Thus, a lower bound is
                 obtained by finding an upper bound for $H(W_t|W^{t-1},\XX^t)$. Because the process $X$
                 changes at each time step, we cannot use the event $A_L$ to obtain an upper bound for
                 $H(W_t|W^{t-1},\XX^t)$ as in the static case. A useful upper bound for $H(W_t|W^{t-1},\XX^t)$ is
                 obtained by using Fano's inequality. Let $P_e$ denote the probability of
                 error in estimating $W_t$ based on observing $Y_t:=(\XX_t,\XX_{t-1}, W_{t-1})$, i.e.,
                 $$P_e = \prob{\hat{W}(Y_t) \neq W_t},$$ where $\hat W(\cdot)$ is a given estimator assumed to be the same for all $t$.
                 Since $W_{t-1}$ is observed, estimating $W_t$ amounts to estimating the increment $N_t = W_t - W_{t-1}$.
                 Because $\XX$ is stationary and $N_t$ is i.i.d., it follows that $P_e$ does not depend on $t$. From Fano's inequality, we have that
                 \begin{eqnarray}\label{eq:lower_bound_dynamic_fano}
                 H(W_t|\XX^t,W^{t-1}) &\leq & H(N_t|Y_t) \nonumber \\
                                      &\leq& H(P_e) + P_e \log_2(1)\nonumber \\
                                                     & = &  H(P_e).
                 \end{eqnarray}
                 Consequently, a lower bound is
                 obtained by combining (\ref{eq:dynamic_lowerB}) with
                 (\ref{eq:lower_bound_dynamic_fano}) above.\footnote{Sharper lower bounds can be obtained by estimating $N_t$ using $(\XX^t,W^{t-1})$.
                 However, the estimate using $Y_t$ is easily computed and already leads to a sharp enough
                 bound.} By letting $t\rightarrow \infty $ we arrive at the following:
                 \begin{theorem} \label{THM:entrpy_dynamic}
                 Consider the vector process $\XX$ consisting of $L$-tuples
                 generated by a Bernoulli random walk with transition probability
                 $p_W$ with $ p_W \leq 1/2$, and the random field $RF=\{X^{(t)}_n : (n,t)\in\mathbb{Z}\times\mathbb{Z}^+, |{\cal X}|<\infty    \}$
                 that is i.i.d. in the $n$ dimension and first-order
                 Markov in the $t$ dimension. The entropy rate of the
                 process $V$ obeys
                 \begin{equation}\label{eq:innovaion_entropy_bd}
                 H(p_W)  +  H(\XX|W)  -H(P_e) \leq H(\XX)  \leq   H(p_W) + H(\XX|W),
                 \end{equation}
                 where $H(\XX|W)$ is as in (\ref{eq:condentropy_rate_innv}), and
                 $P_e$ is the probability of error in estimating $W_1$ based on the observation of $Y_1=(\XX_1,\XX_{0},W_0)$ with any
                 estimator $\hat W(Y_1)$.
                 \end{theorem}
                 The lower and upper bounds become
                 sharp when $P_e\rightarrow 0$. This occurs with large block sizes and for small changes in the
                 background. The examples that follow illustrate
                 the sharpness of the above bounds. In the first
                 example, we consider a binary process $X$, and on
                 the second a Gaussian process with AR(1) temporal
                 innovations.

                 \begin{figure}[htb]
                 \centering
                 \psfrag{BSC}{\small BSC}
                 \psfrag{ldots}{$\ldots$}
                 \psfrag{vdots}{\hspace{0.39cm}$\vdots$}
                 \psfrag{Xt0}{\hspace{-.3cm} \footnotesize $X_{n-1}^{(t)}$}
                 \psfrag{Xt1}{\hspace{-.3cm} \footnotesize $X_{n}^{(t)}$}
                 \psfrag{Xt2}{\hspace{-.3cm} \footnotesize $X_{n+1}^{(t)}$}
                 \psfrag{Yt0}{\hspace{-.3cm} \footnotesize $X_{n-1}^{(t+1)}$}
                 \psfrag{Yt1}{\hspace{-.3cm} \footnotesize $X_{n}^{(t+1)}$}
                 \psfrag{Yt2}{\hspace{-.3cm} \footnotesize $X_{n+1}^{(t+1)}$}
                 \includegraphics[width=7cm]{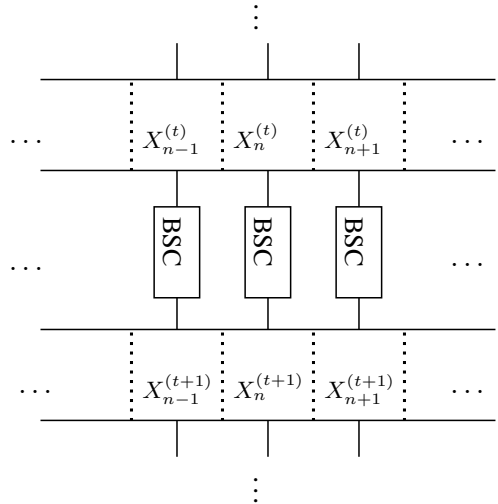}
                 \caption{The binary random field. Innovations are in the form of bit flips caused by binary symmetric channels between
                 consecutive time instants.\label{Fig:BSC_inno}}
                 \end{figure}

\begin{figure}[htb!] \centering
                \subfigure[]{\includegraphics[width=9.5cm]{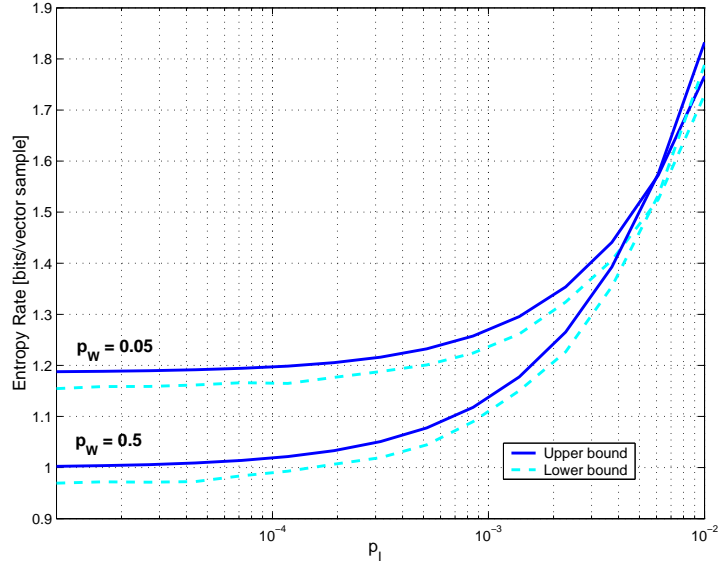}}
                \subfigure[]{\includegraphics[width=9.5cm]{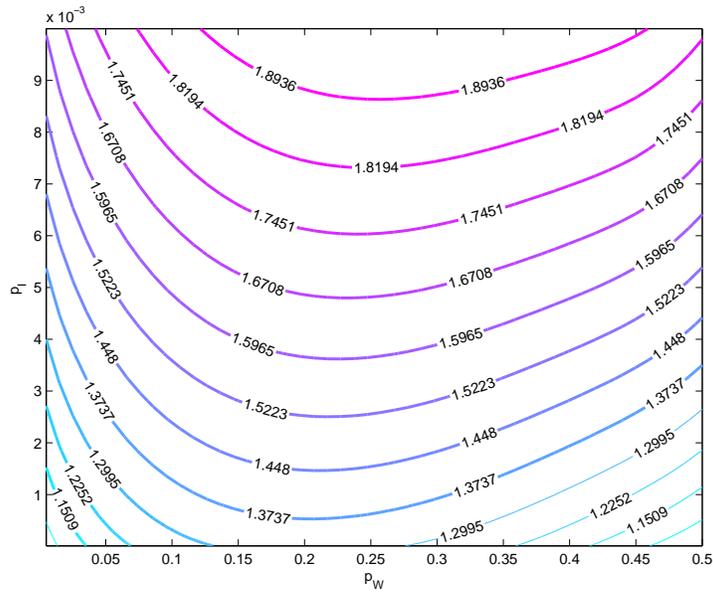}}
                \caption{The binary symmetric innovations. (a) The curves show the lower and upper bounds
                on the entropy rate. Notice that the bounds are sharp for various values of $p_I$. (b) Contour
                plots of the upper bound for various $p_I$ and $p_W$. The lines indicate points of
                similar entropy but with different amounts of spatial and temporal innovation. \label{Fig:Bounds_Innovation_Binary}}
                \end{figure}

\begin{figure}[h!]\centering
                \includegraphics[width = 10cm]{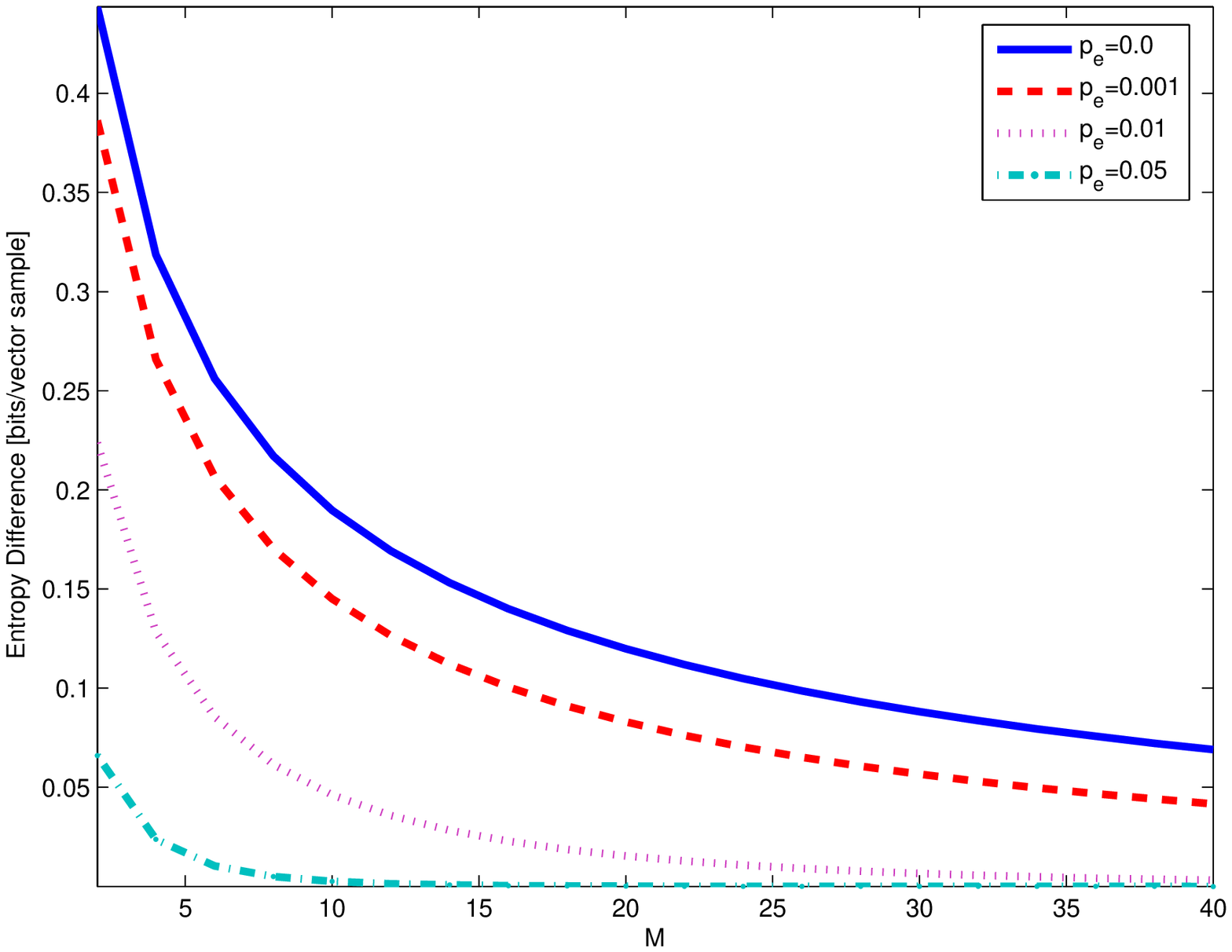}
                \caption{Memory and innovations. Shown is the difference between the
                conditional entropy and the true entropy for the binary innovations with $p_X=0.5$, $p_W=0.5$,
                and $L=8$. The curves show the intuitive fact that when the background changes too rapidly, there is
                little to be gained in bitrate by utilizing more memory. \label{Fig:Memory_Binary_Innovations}}
                \end{figure}

                \begin{example}{\it BSC innovations}\\
                Suppose that at $t=0$, the process is a strip of bits that are i.i.d. Bernoulli
                with initial distribution $p_X$. Suppose that
                from $t$ to $t+1$ there is a nonzero probability
                $p_I$ that the bit $X_n^{(t)}$ is flipped.
                This amounts to a binary symmetric channel (BSC)
                between $X_n^{(t)}$ and $X_n^{(t+1)}$, as illustrated in Figure \ref{Fig:BSC_inno}. The $t$ BSCs
                in series between $X_n^{(0)}$ and $X_n^{(t)}$ are equivalent to a
                single BSC with transition probability (see \cite[p. 221]{CoverThomas:91}, problem 8)
                \begin{equation} \label{eq:channel_equiv}
                p_{I,t} = 0.5 \left (1 - (1-2p_I)^t \right).
                \end{equation}
                Note that for $p_I>0$, we have that $\lim_{t\rightarrow\infty} p_{I,t} = 0.5$. So, for each $n$,
                the distribution of $X_n^{(t)}$ converges to the stationary distribution $Bern(0.5)$.
                Substituting in (\ref{eq:condentropy_rate_innv})
                gives for $p_I>0$:
                \begin{equation} H(V|W) = H(\frac{1}{2}) (1-2p_W) + (L-1)H(p_I)
                +\sum_{i=1}^{ \infty } H(p_{I ,2i})
                \prob{{\retset}_0^{2i}}.
                \end{equation}
                Notice that when $p_I=0$ we recover the static case.
                By using the above in (\ref{eq:innovaion_entropy_bd})
                we obtain the corresponding bounds.
                Figure  \ref{Fig:Bounds_Innovation_Binary} (a)
                illustrates the lower and upper bounds for $L=8$
                and $p_X=0.5$. We compute the bounds using (\ref{eq:condentropy_rate_innv}) and
                (\ref{eq:innovaion_entropy_bd}), where we truncate the infinite sum
                in (\ref{eq:condentropy_rate_innv}) at a very large $t$.
                The probability
                $P_e$ is computed through Monte Carlo simulation using
                a simple Hamming distance detector.
                The bounds are surprisingly robust in this case, and
                provide good approximation  of the true entropy rate.
                Notice that when $p_I$ increases, the entropy rate of the recurrent
                case ($p_W=0.5$) crosses that of the panning case ($p_W=0.05$).
                This is because in the
                recurrent case a greater amount of bits is spent coding the innovations.

                Figure  \ref{Fig:Bounds_Innovation_Binary} (b)
                shows the contour plots of the upper bound for various pairs $(p_I,p_W)$. The plot shows how the two innovations are combined
                to generate a given entropy value. Notice that when $p_W$ approaches $\frac{1}{2}$, the entropy of the trajectory
                 becomes significant and it compensates for the lesser amount of spatial innovation.

                To measure the effect of memory in the dynamic case, we evaluate the upper bound on
                the conditional entropy rate (as in (\ref{eq:4.9})), and the upper bound
                on the true entropy rate given by
                (\ref{THM:entrpy_dynamic}). Figure \ref{Fig:Memory_Binary_Innovations} illustrates the difference between
                the conditional entropy upper bound and the true entropy upper
                bound. The curves
                are similar to the ones obtained in the static case with spatial innovation
                (Figure \ref{Fig:memory}), and confirm the very intuitive fact that memory is less useful when
                the scene around changes rapidly.
                \end{example}

\begin{figure}
                  \centering
                  \includegraphics[width=10cm]{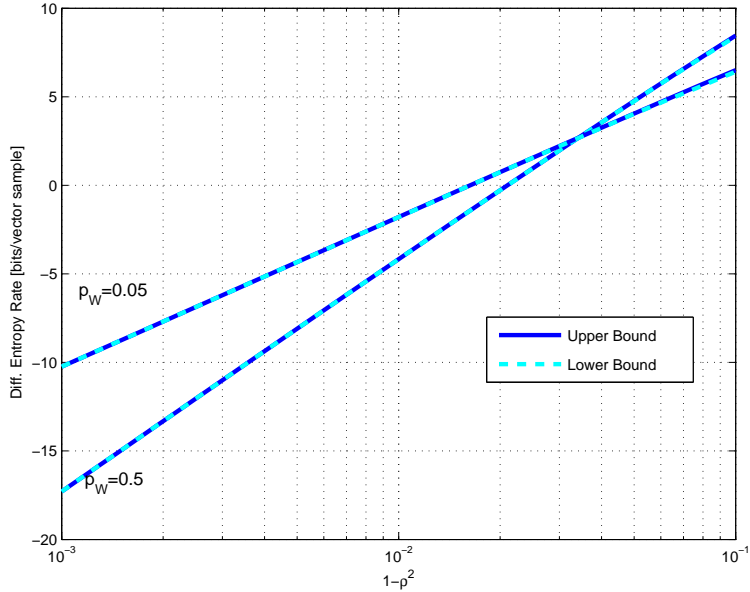}
                  \caption{Differential entropy bounds for the Gaussian AR(1) case as function of the innovation parameter $\rho$.
                  In this example $P_e$ is small enough that the lower and upper bounds
                  practically coincide. Note that the slope of the differential entropy curve is influenced by the value of $p_W$.\label{FIG:Gaussian_Diff_entropy}}
                  \end{figure}

                 \begin{example} AR(1) Innovations.

                 Although the development leading to Theorem \ref{THM:entrpy_dynamic} was made for
                 finite alphabets, the same calculation can be
                 done for a random field taking values on
                 $\mathbb{R}$, provided it has absolutely continuous joint
                 densities. In this case, the entropies involved become differential entropies. For example,
                 for each $n\in\mathbb{Z}$ and $0<\rho<1$, let
                 $$X_n^{(t)}=\rho X_n^{(t-1)} + \epsilon_t$$ for $t\in \mathbb{Z}^+$,
                 where $\epsilon_t \sim N(0,1-\rho^2)$ i.i.d. and independent of $X$.
                 Such a random field model is used for instance in \cite{Hemami:05} for
                 bit allocation over multiple frames. Let $\phi(\sigma^2)$ denote the differential entropy
                  of a Gaussian density with variance $\sigma^2$:

                  $$\phi(\sigma^2):=\frac{1}{2}\log_2(2\pi e \sigma^2).$$

                  It is then easy to check that $h(X_1^{(\infty)}) =
                  \phi(1)$, and $h(X^{(i)}_1|X^{(0)}_1) = \phi(1-\rho^{2 i})$, so that we
                  obtain a lower and an upper bound on the differential entropy rate using Theorem 3. The conditional differential
                  entropy rate $h(V|W)$ is
                  \begin{equation} \label{eq:differential_gaussian}
                  h(V|W) = \phi(1)  (1-2p_W) + (L-1)\phi(1-\rho^2) +\sum_{i=1}^{ \infty }\phi\left(1-\rho^{4i}\right)
                  \prob{{\retset}_0^{2i}}.
                  \end{equation}
                  The infinite sum on the right-hand side is well defined. Because $1-\rho^{2k}$
                  converges to $1$ as $k\rightarrow \infty$ we see that for any value of $\rho$ in $(-1,1)$,
                  the tail of the infinite sum is a sum of positive numbers. Using (\ref{eq:prob_union}) and Lemma \ref{LEMMA:RWALK} (i), we see that
                  $\sum_{i=1}^{ \infty }
                  \prob{{\retset}_0^{2i}} = 2p_W$. Because $\phi(\cdot)$ is
                  concave, we can use Jensen's inequality  as follows:
                  \begin{eqnarray*}
                  \sum_{k=1}^{ \infty } \phi(1-\rho^{4k})
                  \prob{\retset_0^{2k}} & = & 2p_W \sum_{k=1}^{ \infty } \phi(1-\rho^{4k}) \frac{\prob{\retset_0^{2k}}}{2p_W}  \\
                  &\leq  & 2 p_W \phi \left ( \sum_{k=1}^{ \infty } (1-\rho^{4k}) \frac{\prob{\retset_0^{2k}}}{2p_W} \right).
                  \end{eqnarray*}

                  Using Lemma \ref{LEMMA:RWALK} (ii)  and the generating function for the
                  Catalan numbers \cite{KnuthCatalan:89}, one can  further check that
                  $$\sum_{k=1}^{ \infty } (1-\rho^{4k})
                  {\prob{\retset_0^{2k}}}   =  ((1-4(1-p_W)p_W
                  \rho^4))^{1/2} -(1-2 p_W),$$
                  so that the last term is controlled by
                  \begin{equation}
                  \sum_{k=1}^{ \infty }
                  \phi(1-\rho^{4k})\prob{\retset_0^{2k}} \leq 2p_W \phi\left(
                    \frac{(1-4(1-p_W)p_W \rho^4)^{1/2}-(1-2p_W)}{2 p_W}   \right ). \\
                 \end{equation}
                  The above upper bound turns out to be a very good approximation of the infinite sum in (\ref{eq:differential_gaussian}) when $p_W$ is
                  close to $0$, and when $\rho$ is away from $1$.

                  Notice that for $L$ large and $\rho$ close to 1,
                  $P_e$ and $H(P_e)$ are small so that the bounds in Theorem 3
                  are sharp. Figure \ref{FIG:Gaussian_Diff_entropy} displays the bounds on the differential entropy as a function of $\rho$.
                  The bounds are computed
                  following Theorem \ref{THM:entrpy_dynamic} and (\ref{eq:differential_gaussian}).
                  Here $P_e$ is inferred via Monte Carlo simulation with $10^7$ trials, and
                  a minimum MSE detector for $W_t$. The inferred $P_e$ is so low that the lower and upper bounds
                  practically coincide. Analytical computation of $P_e$ is a detection
                  problem beyond the scope of this papers.
                                    \end{example}

                 \subsection{Lossy information rates for the AR(1) random field}

                 Consider the AR(1) innovations of the previous example. Under the MSE distortion measure
                 it is possible to derive an upper bound to the lossy information rate.
                 The key is to compute $R_{\XX^t|W^t}(D)$ defined as in (\ref{eq:conditional_RD_def}) and use the upper bound \cite{Gray:73}:
                 \begin{equation}\label{eq:cond_bound_gray} R_{\XX^t}(D) \leq H(p_W) +
                 R_{V^t|W^t}(D),
                 \end{equation}
                 for each $t>0$. The conditional rate-distortion satisfies the {\it Shannon lower bound} (SLB) \cite{Berger:71}:
                 \begin{equation}\label{eq:SLB_cond}R_{V^t|W^t}(D) \geq  \frac{h(V^t|W^t)}{t}-
                 L\phi(D).\end{equation}
                 The key observation is that for a given fixed trajectory
                 $w^t$, the rate-distortion function of $V^t$ is that of a Gaussian
                 vector consisting of the samples of the random
                 field covered by $W^t$. For a Gaussian vector, the
                 SLB is tight when the per sample distortion
                 is less than the minimum eigenvalue of the
                 covariance matrix (see \cite[p. 111]{Berger:71}). The next
                 proposition gives a condition under which (\ref{eq:SLB_cond}) is tight, and
                 thus when combined with (\ref{eq:cond_bound_gray}) provides an upper bound
                 on the rate-distortion function.

                 \begin{proposition}\label{PROP:SLB_achieve}
                 Consider the vector process $\XX$ resulting from the
                 Gaussian AR(1) random field  with correlation coefficient $0< \rho <1$,
                 and a Bernoulli random walk with probability
                 $p_W\leq 1/2$. The Shannon lower bound for the conditional rate-distortion
                 function is tight whenever the distortion satisfies
                 \begin{equation} \label{eq:SLB_equality_AR(1)}
                 0<D<\frac{1-\rho}{1+\rho}.
                 \end{equation}
                 \end{proposition}

                 To assert the claim we rely on the following lemmas:

                 \begin{lemma}\label{LEMMA:GAUSSIAN_COND_RD}
                 Let $X_1, X_2, \ldots, X_m$ be a sequence of Gaussian vectors in $\mathbb{R}^d$ such that
                 $X_j\sim N(0,C_j)$, and where each $C_j$ has spectrum $\lambda(C_j)$. Let
                 $W$ be a random variable independent of $X_1,\ldots,X_m$ such that
                 $\prob{W=j}=\mu_j$ for $j=1,\ldots,m$.
                 Consider the mixture
                 \begin{equation} X = \sum_{j=1}^m \mathbb{I}_{\{W=j\}}X_j.\end{equation}
                 Denote by $R_{X|W}(D)$ the conditional rate-distortion with per-sample MSE distortion $D$.
                 Then, if \begin{equation} D\leq \min \bigcup_{j=1}^m \lambda(C_j),\end{equation} the conditional rate distortion function is
                 \begin{equation} R_{X|W}(D)=\sum_{j=1}^m \mu_jR_{X_j}(D).\end{equation}
                 \end{lemma}

                 \begin{proof}
                 Let $p(X,\hat X|W)$ be such that ${d}^{-1} \ \mathbb{E}\|X-\hat X\|^2\leq
                 D$. Then,
                 \begin{eqnarray}
                 I(X; \hat X|W) & = &\sum_{j=1}^m \mu_j I(X;\hat X|W=j) \\
                 & \geq & \sum_{j=1}^m \mu_j R_{X_j}(D_j)
                 \end{eqnarray}
                  with \begin{equation} {d}^{-1} \mathbb{E}\|X-\hat X\|^2 = \sum_{j=1}^m \mu_j D_j \leq D,\end{equation} and $D_j := {d}^{-1} \ \mathbb{E}(\|X-\hat X\|^2|W=j)$.
                  The above is minimized when
                  \begin{equation} R_{X_j}'(D_j)=\theta,\end{equation} where $\theta$ is
                  some constant. Suppose $D \leq \min \bigcup_{j=1}^m \lambda(C_j)$ and $D_j=D$. We have
                   \begin{equation} R_{X_j}(D_j)= \frac{1}{d}\sum_{p=1}^d
                   \frac{1}{2}\log_2(\frac{\lambda_{j,p}}{D}),\end{equation}
                   where $\lambda_{j,p}$ are the
                   eigenvalues of $C_j$ and moreover $R'_{X_j}(D_j) = -\frac{1}{2 D_j} = -\frac{1}{2 D} $
                   so that conditions for a minimum are satisfied.
                   The lower bound can be attained by setting
                   $$p(X,\hat X|W=j) =p_j^*(X ,\hat X ),$$ where  $p_j^*(X_j,\hat
                   X_j)$ attains $R_{X_j}(D_j)$.
                  \end{proof}

                 \begin{lemma}(\cite[p. 189]{HornJohnson:99bk}) \label{Lemma:proof_prop_4}
                 Let $\mtx{A}$ be a $n\times n$ Hemitian matrix, and let $1\leq m
                 \leq n$. Let $\mtx{A}_m$ denote a principal submatrix of $\mtx{A}$,
                 obtained by deleting $n-m$ rows and the corresponding columns of
                 $\mtx{A}$. Then, for each integer $k$ such that $1\leq k \leq m$, we
                 have
                 \begin{equation}
                 \lambda_k(\mtx{A}) \leq \lambda_k(\mtx{A}_k),
                 \end{equation}
                 where $\lambda_k(\mtx{A})$ denotes the $k$-th largest eigenvalue of
                 matrix $\mtx{A}$.\\
                 \end{lemma}

                  \noindent {\it Proof of Proposition \ref{PROP:SLB_achieve}}: The SLB for each $t>0$ is given by
                  \begin{equation} R_{V^t|W^t}(D)\geq \label{eq:SLB_proof}\frac{h(\XX^t|W^t)}{t } - L
                  \phi(D).\end{equation}
                  Because $$I(\XX^t;\hat{\XX}^t|W^t) = \sum_{W^t=w^t}
                  \prob{W^t=w^t}I(\XX^t;\hat{\XX}^t|W^t=w^t),$$ in view of Lemma \ref{LEMMA:GAUSSIAN_COND_RD}, it suffices to show that for each
                  $t>0$, and for $0\leq D\leq \frac{1-\rho}{1+\rho}$, the bound
                  $$\frac{I(\XX^t;\hat{\XX}^t|w^t)}{t} \geq \frac{h(\XX^t|w^t)}{t  } - L \phi(D), \text{ for }
                  \mathbb{E}(d(\XX^t,\hat \XX^t)|w^t) \leq D $$ is achievable.
                  Given $W^t=w^t$, the above bound is attainable if $D$ is
                  smaller than the minimum eigenvalue of the covariance
                  matrix of the random field samples covered by
                  $w^t$. Denote this covariance by $\mtx{C}_{w^t}:=\text{Cov}(V^t|w^t)$. Because the random field is
                  independent in the spatial dimension $n$, the spectrum of the
                  covariance matrix is the disjoint union of the spectra of the covariance
                  matrices corresponding to the random field samples of $V^t$ at similar location $n$. Each
                  $\mtx{C}_{w^t}$ is a submatrix
                  of the $t\times t$ Toeplitz matrix $\mtx{T}_t(\rho)$ with
                  entries $[\mtx{T}_t(\rho)]_{ij}=\rho^{|i-j|}$. Since $\lambda_{min}(\mtx{T}_t(\rho))$
                  decreases to $(1-\rho)/(1+\rho)$ as $t\rightarrow \infty$ \cite{GrayToeplitz:72},  by
                  applying the Lemma above we
                  conclude that
                  \begin{equation}\lambda_{min}(\mtx{C}_{w^t})  \geq
                  \lambda_{min}(\mtx{T}_t(\rho))  \geq  \frac{1-\rho}{1+\rho}.
                  \end{equation}
                  Therefore, the bound (\ref{eq:SLB_proof}) is achievable for each $t$ and since
                  the limit of $R_{\XX^t|W^t}(D)$ exists it follows
                  that  the bound is achievable for $t\rightarrow
                  \infty$.\endproof


%
%

\begin{figure}[h!]\centering
                  \subfigure[]{\includegraphics[width = 7.2cm]{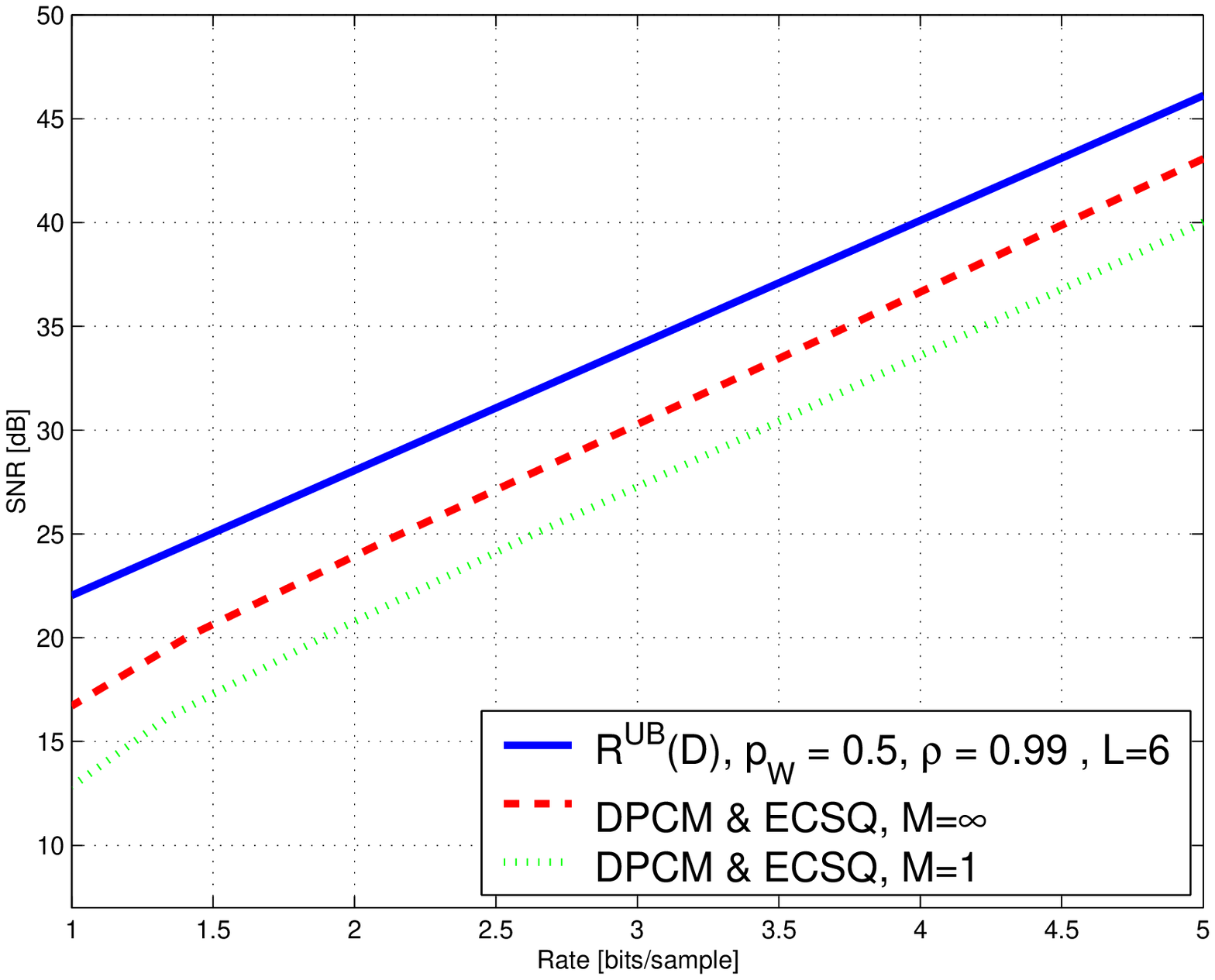}} \
                  \subfigure[]{\includegraphics[width = 7.2cm]{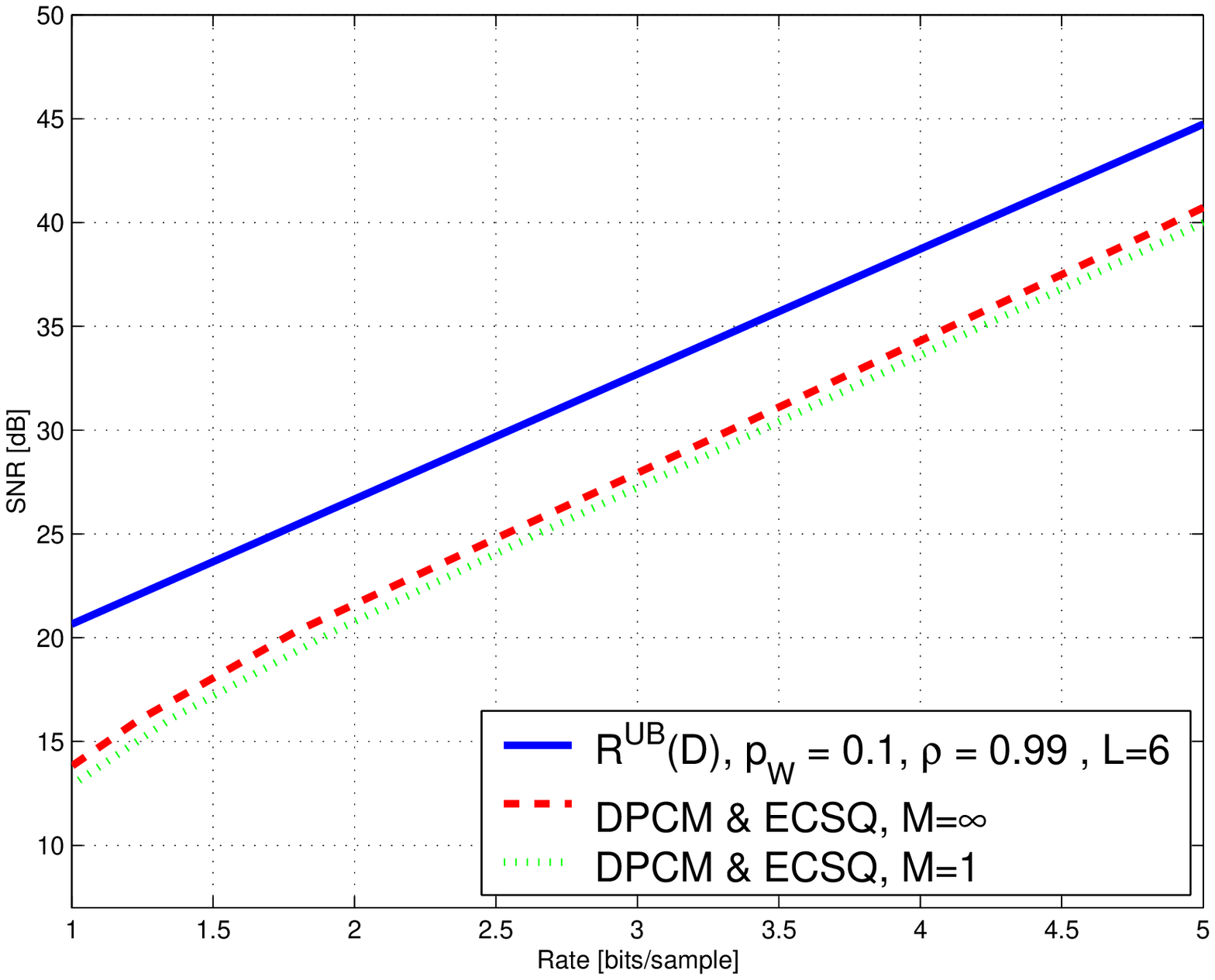}} \
                  \subfigure[]{\includegraphics[width = 7.2cm]{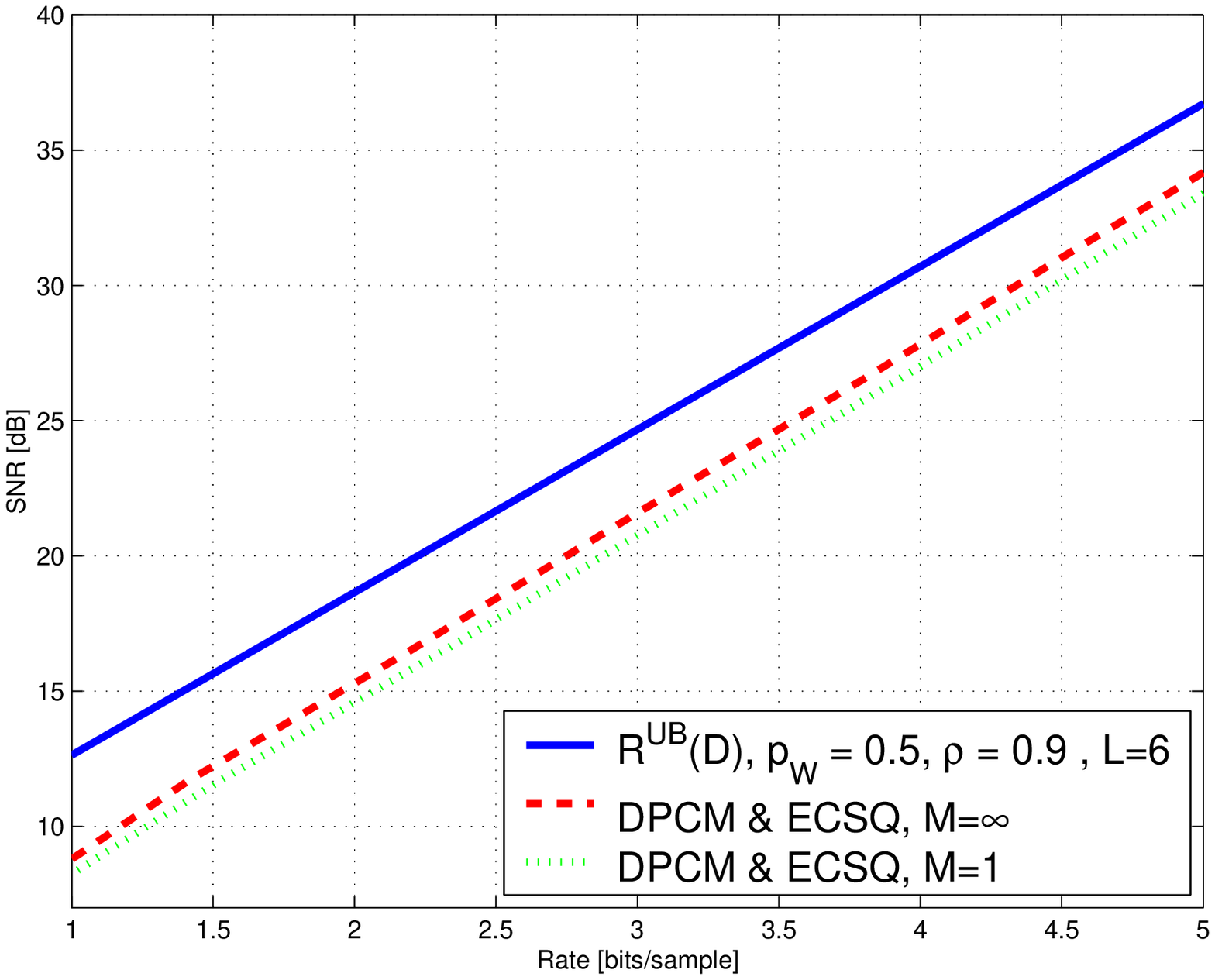}} \

                  \caption{Performance of DPCM with motion for various
                  $\rho$ and $p_W$. For $\rho=0.99$ and $\rho = 0.9$
                  the upper bound is valid for SNR greater than 23 dB and 12.8 dB, respectively.
                  (a) Memory provide considerable gains, $p_W=0.5, \rho = 0.99$. (b) Modest
                  gains when $p_W=0.1$. (c) Modest gains when $\rho = 0.9$, as background changes too rapidly. \label{Fig:DPCM}}
                  \end{figure}

                  \begin{example}

                  We simulate the AR(1) dynamic reality model.
                  To compress
                  the process $\XX^t$, we estimate the trajectory and send it as side information. With the
                  trajectory at hand, we encode the samples with DPCM, encoding the residual with entropy constrained
                  scalar quantization (ECSQ). We build two encoders. In the first one, prediction is done
                  utilizing only the previously encoded vector sample; in the second, all encoded samples up to time $t$
                  are available to the encoder (and decoder).
                  Figure  \ref{Fig:DPCM} illustrates the SNR as a function of rate when the block-length $L=8$.
                  In Figure  \ref{Fig:DPCM} (a) and (b) we have $\rho = 0.99$ and the upper bound is valid for SNR
                  greater than 23 dB. In Figure \ref{Fig:DPCM} (a), we have $p_W=0.5$. Because the scene changes slowly and is highly
                  recurrent, the infinite memory encoder ($M=\infty$) is about 3.5 dB better than when
                  $M=1$. The same behavior is not observed when the scene is not recurrent
                  (panning case, $p_W=0.1$, Figure  \ref{Fig:DPCM} (b) ),
                   and when the background changes too rapidly ($\rho = 0.9$, Figure  \ref{Fig:DPCM}
                   (c)).%
%

                   \end{example}

\section{Conclusion} \label{Sec:Conc}

We have proposed a stochastic model for the plenoptic function
that enables the precise computation of information rates. For the
static case, we provided lossless and lossy information rate
bounds that are tight in a number of interesting cases. In some
scenarios, the theoretical results support the ubiquitous hybrid
coding paradigm of extracting motion and coding a motion
compensated sequence.

We extended the model to account for changes in the background
scene, and computed bounds for the lossless and lossy information
rates for the particular case of AR(1) innovations. The bounds for
this ``dynamic reality'' are tight in some scenarios, namely when
the background scene changes slowly with time (i.e., $\rho$ close
to 1).

The model explains precisely how long-term motion prediction helps
coding in both static and dynamic cases. In the dynamic model,
this is related to the two parameters $(p_W, \rho)$ that
symbolizes the rate of recurrence in motion and the rate of
changes in the scene. As $(p_W, \rho) \rightarrow (0.5,1)$, long
term memory predictions result in significant improvements (in
excess of 3.5 dB). By contrast, if either $\rho$ is away from 1,
or if $p_W$ is away from $0.5$, long term memory brings little
improvement.

Although we developed the results for the Bernoulli random walk,
the model can be generalized to other random walks on $\mathbb{Z}$
and $\mathbb{Z}^2$. Our current work includes such
generalizations. It also includes estimating $\rho$ and $p_W$ for
real video signals and fitting the model to such signals.

\bibliographystyle{IEEEtran}
\bibliography{INFO_RATE}

\end{document}